\documentclass[aps,preprint]{revtex4}
\usepackage[dvipdfm]{graphicx}
\graphicspath{{figs/}}
\usepackage{amsmath,amsthm,amssymb}

\newcommand{\rs}[1]{_{\rm #1}}             
\newcommand{\bm}[1]{\mbox{\boldmath $#1$}} 

\newcommand{\noscmax}{N_{\rm osc}^{\rm max}} 

\begin{document}

\title{
     Efficient method to perform quantum number projection
     and configuration mixing
     for most general mean-field states
}

\author{Shingo Tagami and Yoshifumi R. Shimizu}

\affiliation{Department of Physics, Graduate School of Science,
Kyushu University, Fukuoka 812-8581, Japan}


\begin{abstract}
Combining several techniques,
we propose an efficient and numerically reliable
method to perform the quantum number projection and configuration mixing
for most general mean-field states,
i.e., the Hartree-Fock-Bogoliubov (HFB) type product states
without symmetry restrictions.
As for example of calculations, we show the results of
the simultaneous parity, number and angular-momentum projection
from HFB type states generated from the cranked Woods-Saxon mean-field
with a very large basis that is composed of $N\rs{max}=20$
spherical harmonic oscillator shells.
\end{abstract}

\maketitle

\section{Introduction}
\label{sec:intro}

Recent advent of radioactive beam facilities is extending more and more widely
the research area of nuclear physics.
It is increasingly important to have a unified understanding of
nuclear structure in various regions of the nuclear chart,
with variety of ingredients such as
shell effects, deformations and collective motions
like rotation and vibration.
Undoubtedly, the basic starting point is the selfconsistent mean-field
approximation~\cite{RS80,BR85}, e.g., the Hartree-Fock (HF) or
the Hartree-Fock-Bogoliubov (HFB) method including pairing correlations,
with suitably chosen density-dependent effective interactions,
or in a modern terminology, energy density functionals;
see e.g. Ref.~\cite{BHR03}.
With the symmetry-breaking, relatively simple mean-field states
can take into account most of the many-body correlations
in a very efficient way~\cite{RS80,BR85},
and have been successfully applied to study various nuclear
phenomena not only near the ground state but also
in the low- to high-spin excited states~\cite{AFN90,Fra01,SW05}.

However, the symmetry-breaking mean-field description is not enough
because it represents merely the intrinsic state and the broken symmetry
should be recovered in the laboratory frame.
One of the most important consequences of the symmetry-breaking
is the appearance of the symmetry-restoring collective motion.
One well-adopted method to restore the symmetry is to utilize
a phenomenological collective model like the rotor description
in the unified model of Bohr-Mottelson~\cite{BM75}.
One can consistently introduce the redundant collective coordinate
into the many-body theory by employing the quantum mechanical
constraint formalism (the gauge theory)
between the nucleon and collective degrees of freedom,
see e.g. Ref.~\cite{BK90}; its exact treatment is rather involved.
Another way, without recourse to the external variables,
to restore the symmetry within the nucleon degrees of freedom is
the quantum number projection~\cite{RS80,BR85}.
Existence of a symmetry-breaking mean-field state
means that all the states connected by
the symmetry operation, e.g., rotation of the system, are degenerate.
Superposition of all these degenerate states gives a better
quantum mechanical description in the variational sense,
as it is clear in the formulation of the generator coordinate method (GCM).
Since the symmetry requires the specific form of weight functions
for superposition, the procedure restores the symmetry at the same time,
i.e., projects out the states with good quantum numbers.

The most important symmetry-breaking in nuclear structure is
the spatial deformation, e.g., the quadrupole shape, so that
the projection of the angular momentum is necessary to obtain
the eigenstates of the angular momentum operators, especially for calculating
the electromagnetic transition probabilities.
There is a long history in the angular momentum projection calculations.
Except for some special calculations intended for very light nuclei,
the general framework of the projection from the (HFB-like)
general product-type mean-field wave functions
has been developed by K.~Hara and his collaborators
in Refs.~\cite{HI79,HIT79,HI80}, where the calculation is restricted
to the axially symmetric shape, but extended to the triaxially deformed
and cranked (for high-spins) cases in Ref.~\cite{HHR82}
(see also the review paper~\cite{HS95},
and a more recent application~\cite{ETY99}).
Based on the angular momentum projection,
the variation after projection calculations from general mean-field states
have also been performed for the $G$-matrix based realistic interactions,
see e.g. Refs.~\cite{SG87,Sch04}.
However, relatively small model spaces are used in these works,
e.g., the two or three harmonic oscillator shells.
Recently, the angular momentum projection with much larger space has been
attempted with restriction of axially symmetry,
intending to employ the Skyrme (or more general)
energy functional~\cite{VHB00}, where the GCM calculation
with respect to the quadrupole deformed coordinates on top of it
is performed (see also Ref.~\cite{RER02} for the similar type
calculation with the finite range Gogny interaction).
The restriction of axial symmetry has been lifted
in more recent works for the Skyrme~\cite{BH08},
the relativistic mean-field~\cite{YM09,YM10},
and the Gogny~\cite{RE10} approaches,
although still the time-reversal invariance (no-cranking)
and the $D_2$ symmetry of deformation are imposed in such calculations.

In this paper, we discuss an efficient method of general quantum number
projection, i.e., rather technical aspect of projection.
We intend to perform the angular momentum projection with
other projections, the number and parity, at the same time from
the most general symmetry-breaking HFB type mean-field, i.e.,
the axial symmetry, the parity, as well as the time-reversal
invariance are broken.  For this kind of the most general projection,
the frequently used speed-up technique, for example,
using the $D_2$ symmetry
that reduces the integration volume of the three Euler angles by factor 16,
cannot be utilized.
Therefore an efficient method to perform the projection is crucial.
The basic ingredient of the projection is the overlap of operators
between the product-type mean-field wave functions, based on
the generalized Wick theorem~\cite{OY66,BB69}.  For evaluation of
such overlaps many matrix operations composed of the multiplication,
inverse and determinant, are necessary, and so the dimension of matrices
is the most crucial factor.

One of the essential ideas of our efficient method has been
invented and discussed already in the Appendices of Ref.~\cite{BFH90};
in fact we have noticed this reference
after finishing our work (see also Refs.~\cite{Rob94,VHB00,YM09}).
It is based on the fact that, although the model space for calculating
the realistic single-particle states are large, the effective number
of states contributing the HFB type product states are relatively small
because the nuclear superfluidity is not so strong:
If there is no pairing correlation, the mean-field state is
a Slater determinant composed of the single-particle wave functions
whose number is nothing else but the number of constituent particles.
The truncation of the effective model space reduces
the dimension of matrices dramatically
in the most essential part of the calculation.
Another important point of our method is that we make full use of
the Thouless amplitudes rather than the $(U,V)$ amplitudes of
the generalized Bogoliubov transformation~\cite{RS80}.
One of the reasons for this is that the sign of the norm overlap between
general product-type wave functions can be precisely calculated
by using their Thouless amplitudes~\cite{Rob09}.
Moreover, as a vacuum state, with respect to which the Thouless form
of the general HFB type state is considered, we employ a suitably
chosen Slater determinantal state, i.e., the particle-hole vacuum
in place of the true nucleon vacuum is utilized.
This increases the numerical stability
of the Thouless amplitude ($Z=(VU^{-1})^*$) for the case
of vanishing pairing correlations on one hand, and makes it possible
to truncate further the space composed of deep hole states
(the core contributions) on the other hand,
which is very effective for calculations of heavy nuclei.

This article is organized as follows.
The basic formulation of the efficient method to
perform the general quantum number projection
and/or the configuration mixing is presented
in Sec.~\ref{sec:formulation}.
The results of example calculations are shown and discussed
in Sec.~\ref{sec:example}.
Sec.~\ref{sec:summary} is devoted to the summary.

\section{Formulation}
\label{sec:formulation}

\subsection{Norm overlap, contractions, and generalized Wick theorem}
\label{sec:Wick}

Although the basic method of the projection (or GCM) is
well-known~\cite{HI79,RS80}, we recapitulate it in order to fix
the notation and explain our specific treatment
(we mainly follow the notation of Ref.~\cite{RS80}).

The Hamiltonian is composed of the one-body part and the two-body interaction,
\begin{equation}
 \hat H = \sum_{l_1 l_2} t_{l_1 l_2} \hat c_{l_1}^\dagger \hat c_{l_2}^{}
  +\frac{1}{2}  \sum_{l_1 l_2 l_3 l_4} v_{l_1 l_2 l_3 l_4}
  \hat c_{l_1}^\dagger \hat c_{l_2}^\dagger \hat c_{l_4}^{} \hat c_{l_3}^{},
\label{eq:Ham2bd}
\end{equation}
whose explicit form is specified in the next section.
Here $(\hat c_l^\dagger, \hat c_l^{})$ ($l=1,2,...,M$) are
the basic particle (nucleon) creation and annihilation operators,
with $M$ being the number of basis states.
The appropriate choice of this original basis is very important
to perform the angular momentum projection effectively.
In this paper we choose the spherical (isotropic) harmonic oscillator basis,
$\{|Nljm\rangle \}$, where the selection of the harmonic oscillator
is optional; what is important is that the angular momentum $(jm)$ is
a good quantum number because the representation of the rotation
matrix in this basis is block diagonal.
Another possible choice for the case that the two-body interaction
is local like the Skyrme and Gogny forces is the isotropic Cartesian
harmonic oscillator basis, in which the rotation operator about
one of the three axes is again represented by a block diagonal matrix
with very few non-zero elements.

The fundamental object for the projection (or GCM) calculation
is overlap $\langle\Phi|\hat O|\Phi'\rangle$
of an arbitrary operator $\hat O$ between two general
product-type states $|\Phi\rangle$ and $|\Phi'\rangle$.
These mean-field states are vacuums of the quasiparticle operators,
$\hat \beta_k$ and $\hat \beta_k'$, respectively, which are related
to the original particle basis by the following
general Bogoliubov transformations,
\begin{equation}
 \hat \beta_{k}^\dagger = \sum_{l} \left[
  U_{lk}\hat c_{l}^\dagger + V_{lk}\hat c_{l}^{}
 \right] ,\qquad
 \hat \beta_{k}'^\dagger = \sum_{l} \left[
  U'_{lk}\hat c_{l}^\dagger +V'_{lk}\hat c_{l}^{}
 \right] .
\label{eq:GBtra}
\end{equation}
In most of realistic situations these $(U,V)$ amplitudes are
provided, although they contains redundant degree of unitary
transformations between the quasiparticles for uniquely
specifying the HFB type state~\cite{RS80}.
In this subsection we assume that the states
$|\Phi\rangle$ and $|\Phi'\rangle$ are not orthogonal
to the true vacuum $|\rangle$ of the original particle $\hat c_l^{}$.
Employing the Thouless theorem, these quasiparticle vacuums can be
written explicitly as
\begin{equation}
\begin{array}{ll}
 |\Phi \rangle = n \, e^{\hat Z} |\rangle ,& \quad
 {\displaystyle
 \hat Z \ \equiv \frac{1}{2} \sum_{l'l} Z_{l'l}
   \hat c_{l'}^\dagger \hat c_{l}^\dagger},\cr
 |\Phi' \rangle = n' \, e^{\hat Z'} |\rangle ,& \quad
 {\displaystyle
 \hat Z' \ \equiv \frac{1}{2} \sum_{l'l} Z'_ {l'l}
   \hat c_{l'}^\dagger \hat c_{l}^\dagger},
\end{array}
\label{eq:ThoulZ}
\end{equation}
with normalization constants $n\equiv \langle |\Phi\rangle$ and
$n'\equiv \langle |\Phi'\rangle$, and
the Thouless amplitudes $Z$ and $Z'$ are defined
in an obvious matrix notation by
\begin{equation}
 Z \equiv \left(V U^{-1}\right)^* , \qquad
 Z' \equiv \left(V' U'^{-1}\right)^*.
\label{eq:Zdef}
\end{equation}
In the following, we use the conventional matrix notations,
$A^\dagger$ (Hermitian conjugate), $A^T$ (transpose),
$A^*$ (complex conjugate) and $A^{-1}$ (matrix inverse),
and further $A^{-\dagger}\equiv (A^\dagger)^{-1}$,
$A^{-T}\equiv (A^T)^{-1}$ and $A^{-*}\equiv (A^*)^{-1}$.
Then, the norm overlap is given by
\begin{equation}
 \langle\Phi|\Phi'\rangle
 = n^* n'\,
 \left(\det\left[1+Z^\dagger Z' \right]\right)^{1/2}
 = n^* n'\, (-1)^{M(M+1)/2}
  \, {\rm pf} \begin{pmatrix}
  Z' & -1 \cr 1 & Z^\dagger \end{pmatrix},
\label{eq:normovl}
\end{equation}
where the sign of square root of the determinant is uniquely fixed
by the calculation of the pfaffian~\cite{Rob09}.
If we impose the normalization condition,
$\langle\Phi|\Phi \rangle=1$ and $\langle\Phi'|\Phi'\rangle=1$,
the absolute value $|n|$ and $|n'|$ are determined.
Using the identity
$|\det{U}|^2=\det{UU^\dagger}=\left(\det\left[1+Z^\dagger Z\right]\right)^{-1}$,
we may write
\begin{equation}
 n=e^{i\theta}\left(\det U^* \right)^{1/2},\qquad
 n'=e^{i\theta'}\left(\det U'^* \right)^{1/2},
\label{eq:normphase}
\end{equation}
where the quantities $e^{i\theta}$ and $e^{i\theta'}$ fix the 
phases of the states $|\Phi\rangle$ and $|\Phi'\rangle$, respectively.

The overlap of an arbitrary operator is calculated according to
the generalized Wick theorem~\cite{OY66,BB69}; for example,
for the two-body interaction,
\begin{equation}
 \frac{ \langle\Phi | \hat c_{l_1}^\dagger \hat c_{l_2}^\dagger
  \hat c_{l_4}^{} \hat c_{l_3}^{} |\Phi' \rangle}
 { \langle \Phi|\Phi' \rangle } =
  \rho^{(c)}_{l_3 l_1} \rho^{(c)}_{l_4 l_2}
 -\rho^{(c)}_{l_4 l_1} \rho^{(c)}_{l_3 l_2}
 +\bar\kappa^{(c)}_{l_2 l_1} \kappa^{(c)}_{l_3 l_4},
\label{eq:ctwobody}
\end{equation}
where the basic contractions, or the transition density matrix
$\rho^{(c)}$ and
the transition pairing tensors, $\kappa^{(c)}$ and $\bar\kappa^{(c)}$
with respect to the original particle basis ($\hat c^\dagger_{},\hat c^{}_{}$)
are defined by
\begin{equation}
\begin{array}{ll}
 \rho^{(c)}_{l'l} \equiv&
 {\displaystyle
 \frac{ \langle \Phi|\hat c_{l}^\dagger \hat c_{l'}^{}|\Phi' \rangle }
 { \langle \Phi|\Phi' \rangle }
 = \left(Z' \left[ 1+Z^\dagger Z' \right]^{-1} Z^\dagger\right)_{l'l},
 }\cr
 \kappa^{(c)}_{l'l} \equiv&
 {\displaystyle
 \frac{ \langle \Phi|\hat c_{l}^{} \hat c_{l'}^{}|\Phi' \rangle }
 { \langle \Phi|\Phi' \rangle }
 = \left(Z' \left[ 1+Z^\dagger Z' \right]^{-1}\right)_{l'l},
 }\cr
 \bar\kappa^{(c)}_{l'l} \equiv&
 {\displaystyle
 \frac{ \langle \Phi|\hat c_{l}^\dagger \hat c_{l'}^\dagger|\Phi' \rangle }
 { \langle \Phi|\Phi' \rangle }
 = \left(\left[ 1+Z^\dagger Z' \right]^{-1} Z^\dagger\right)_{l'l}.
 }
\label{eq:contc}
\end{array}
\end{equation}

In Ref.~\cite{HI79}, for example, the contractions between the quasiparticles
$\hat \beta_k$ and $\hat \beta_k'$ are given in terms of the coefficients
of generalized Bogoliubov transformation between them.
However, it is shown in the following subsections that the truncation of
the effective model space can be done in a more transparent manner
if the Thouless amplitudes are utilized.

\subsection{Quantum number projection}
\label{sec:proj}

A state with good quantum numbers is obtained by the projection
from the symmetry-breaking mean-field state $|\Phi\rangle$~\cite{RS80},
\begin{equation}
 |\alpha\rangle=\hat P_\alpha |\Phi\rangle,\qquad
 \hat P_\alpha = \int g_\alpha(\bm{x}){\hat D}(\bm{x})d\bm{x}.
\label{eq:ProjD}
\end{equation}
Here $\alpha$ denotes a set of quantum numbers,
and the projection operator $\hat P_\alpha$ is defined
by the superposition of all possible unitary transformations
$\hat D(\bm{x})$ with the weight function $g_\alpha(\bm{x})$,
where the continuous parameters $\bm{x}\equiv (x_1,x_2,...)$ specify
the coordinates in the manifold of symmetry operations.
In the case of the number and the angular momentum projection,
it is written as $\hat D(\bm{x})=e^{i\varphi \hat N} \hat R(\omega)$
with the gauge angle $\varphi$ and the Euler angles $\omega$
as parameters $\bm{x}$, where
$\hat N$ is the number operator and $\hat R(\omega)$
is the rotation operator.  Note that the parity projector,
\begin{equation}
 \hat P_\pm = \frac{1}{2}\left[1 \pm \hat \Pi \right],
\label{eq:ProjP}
\end{equation}
where $\hat \Pi$ is the space inversion operator,
has the same form as in Eq.~(\ref{eq:ProjD}),
although the values of parameter are discrete.

General quantum-number-projection calculation requires
to evaluate the matrix elements,
$\langle \Phi | {\hat P}_\alpha {\hat O} {\hat P}_{\alpha'}| \Phi' \rangle$,
between two general product-type states $|\Phi \rangle$ and $|\Phi' \rangle$
for arbitrary operator $\hat O$.  Since the operator $\hat O$,
e.g., the Hamiltonian or the electromagnetic transition operators,
usually belongs to an irreducible representation of the symmetry transformation
$\hat D(\bm{x})$ associated with the projector,
it is enough to consider
either $\langle \Phi | {\hat O} {\hat D}(\bm{x})| \Phi' \rangle$
or $\langle \Phi | {\hat D}(\bm{x}) {\hat O}| \Phi' \rangle$
at mesh points of numerical integration over the parameter space $(\bm{x})$.
We employ the form where the unitary
transformation is on the right in the following,
but one can use another form with trivial modifications.

In the following we omit to denote the parameters $\bm{x}$
in the unitary transformation $\hat D$
as long as there is no confusion.
In the usual projection calculations, the unitary transformation
is generated by a one-body Hermitian operator $\hat G$
($\hat G^\dagger=\hat G$); most generally,
\begin{equation}
 \hat D=e^{i\hat G},\qquad
 \hat G = g^{0}+\sum_{ll'} g^{11}_{ll'} \hat c_{l}^\dagger \hat c_{l'}^{}
 +\frac{1}{2}\sum_{ll'}\left(g^{20}_{ll'} \hat c_{l}^\dagger \hat c_{l'}^\dagger
  +\mbox{h.c.}\right).
\label{eq:DandGgen}
\end{equation}
The norm overlap of two normalized HFB type states
$|\Phi\rangle$ and $|\Phi'\rangle$ in the Thouless form
is given in Eq.~(\ref{eq:normovl})
with the normalization constants in Eq.~(\ref{eq:normphase}).
If the one state is the unitary transformed state of the other,
$|\Phi'\rangle=\hat D|\Phi\rangle$,
the relative phase between them is determined uniquely~\cite{HI79}.
Namely, the difference between $\theta'$ and $\theta$
in Eq.~(\ref{eq:normphase}) in such a case is given by
\begin{equation}
 \theta'-\theta=g^0+\frac{1}{2}{\rm Tr}\,g^{11}
 \equiv \Theta(\hat D),
\label{eq:Dphase}
\end{equation}
and then the norm overlap can be calculated as
\begin{equation}
 \langle\Phi|\hat D|\Phi\rangle=
 e^{i\Theta(\hat D)}
 \left(\det {\cal U}^*\right)^{1/2},\qquad
 {\cal U}= U^\dagger U'+V^\dagger V' .
\label{eq:Dmean}
\end{equation}

\subsection{Model space truncation}
\label{sec:trunc}

The occupation probabilities of the original particle basis are
not necessarily small for a given quasiparticle vacuum
state $|\Phi\rangle$,
\begin{equation}
 \langle \Phi|\hat c_{l}^\dagger \hat c_{l}^{}|\Phi\rangle
 \ne 0, \quad l=1,2,...,M.
\end{equation}
However, the superfluidity of nuclei is not so strong in most cases
and the effective number of basis states contributing to the state
$|\Phi\rangle$ is relatively small:  It can be clearly recognized
by introducing a canonical-like basis
that diagonalizes the usual density matrix
$\rho_{l'l}\equiv
 { \langle \Phi|\hat c_{l}^\dagger \hat c_{l'}^{}|\Phi \rangle }
 /{ \langle \Phi|\Phi \rangle }$;
\begin{equation}
 \hat b_k^\dagger = \sum_l W_{lk} \hat c_l^\dagger , \quad
  WW^\dagger = W^\dagger W=1,
\label{eq:canb}
\end{equation}
\begin{equation}
 \rho = W \bar\rho W^\dagger, \quad
 \bar\rho = \mbox{diag}(v_1^2,v_2^2,...),
\label{eq:Wcanb}
\end{equation}
where the occupation probabilities
$v_k^2=\langle \Phi|\hat b_{k}^\dagger \hat b_{k}^{}|\Phi\rangle$
$(k=1,2,...,M)$,
which are at least pairwisely degenerate,
are assumed to be in descending order
(i.e., $v_1^2=v_2^2 \ge v_3^2=v_4^2 \ge...$).
Then most of $v_k$'s are negligibly small; more precisely, we take
some small number $\epsilon$ and select the $P$ space composed
of $L_p(\epsilon)$ orbits which satisfy $v_k^2 \ge\epsilon$
($k=1,2,...,L_p(\epsilon))$, while in the complemental $Q\,(=1-P)$ space
we set $v_k=0$ ($k=L_p(\epsilon)+1,...,M$).
Practically the parameter $\epsilon$ is chosen to be as large as
possible within the condition that the final results (e.g., the energy spectra)
do not change; we find that typically $\epsilon=10^{-4}-10^{-5}$ is enough.
For example, if we use a Woods-Saxon potential,
the typical number of spherical oscillator shells necessary is
$N_{\rm osc}\approx 12-14$ for heavy stable nuclei.
However, it can happen that one should include more shells,
e.g., up to $N_{\rm osc}\approx 20$,
to describe weakly bound orbits correctly, then $M > 3000$.
It turns out that the effective number of the $P$ space stays
$L_p(\epsilon)\approx 100-250$ in most of the cases (for either
neutrons or protons), which are about one order of magnitude smaller
than the number of original basis states $M$.
In Ref.~\cite{Taj04}, this fact is used and a very efficient method
is developed to solve the HFB equation in terms of the small number
of canonical basis (note that the canonical basis is usually calculated
after obtaining the HFB state).

Employing the Block-Messiah theorem,
the amplitudes of the Bogoliubov transformation in Eq.~(\ref{eq:GBtra})
is written as
\begin{equation}
 U = W \bar U C ,\quad V = W^* \bar V C,
\label{eq:canWU}
\end{equation}
where the matrices $W$ (the one in Eq.~(\ref{eq:canb})) and $C$ are unitary,
and $(\bar U,\bar V)$ are of the so-called canonical form~\cite{RS80}
if the basis is rigorously canonical,
which is not necessarily required in the following discussion.
According to the $P$ and $Q$ space decomposition defined above,
they are in the following block forms,
\begin{equation}
 W = \begin{pmatrix}
  W_p & W_q \end{pmatrix} , \quad
 \bar U = \begin{pmatrix}
  \bar U_{pp} & 0 \cr 0 & 1 \end{pmatrix} , \quad
 \bar V = \begin{pmatrix}
  \bar V_{pp} & 0 \cr 0 & 0 \end{pmatrix}.
\label{eq:WUVpq}
\end{equation}
where obviously, for example, $W_p$ is $M\times L_p$ matrix and
$\bar U_{pp}$ is $L_p\times L_p$ matrix
(dropping $\epsilon$ for simplicity).
Although the effective number of the $P$ space ($L_p$) is relatively small,
it should be noted that calculations of the projection,
especially the angular momentum projection,
are not confined within this space.
This is because of the symmetry-breaking feature
of the general quasiparticle state $|\Phi\rangle$;
the transformation in the projection operation (e.g. the rotation)
kicks the orbits belonging to the $P$ space out of the model space.

For the number or angular momentum projection,
the one-body generator $\hat G$ of the symmetry transformation $\hat D$
in Eq.~(\ref{eq:DandGgen}) has no $g^{20}$ terms in the original basis;
\begin{equation}
 \hat G = g^{0}+\sum_{ll'} g^{11}_{ll'} \hat c_{l}^\dagger \hat c_{l'}^{},
\label{eq:DandG}
\end{equation}
and then
the $M\times M$ transformation matrix $D$ in the original basis $\hat c_l^{}$
is defined by
\begin{equation}
 \hat D \hat c_l^\dagger \hat D^\dagger
 = \sum_{l'} D_{l'l} \hat c_{l'}^\dagger ,\qquad
  D = \exp (i g^{11}).
\label{eq:Dgmat}
\end{equation}
Then, assuming the normalization, $\langle \Phi|\Phi\rangle=1$,
and using the identity $\exp{(i{\rm Tr}\,g^{11})}=\det{D}$,
the norm overlap in Eq.~(\ref{eq:Dmean}) is explicitly written as
\begin{equation}
 \langle \Phi|\hat D|\Phi \rangle
 =e^{ig^0} \Bigl( \det{D}\det{\cal U}^* \Bigr)^{1/2}
 =e^{ig^0} \Bigl( \det{\tilde D}\det{\bar {\cal U}}^* \Bigr)^{1/2},
\label{eq:DUovlp}
\end{equation}
with
\begin{equation}
 {\cal U} = U^\dagger D U +V^\dagger D^* V
  = C^\dagger \bar {\cal U} C, \qquad
 \bar {\cal U} = \bar U^\dagger \tilde D \bar U
 +\bar V^\dagger \tilde D^* \bar V.
\end{equation}
The matrix $\tilde D$ is the transformation matrix
in the canonical basis $(\hat b^\dagger_{},\hat b^{}_{})$,
\begin{equation}
 \tilde D  \equiv \ W^\dagger D W =
 \begin{pmatrix}
 W_p^\dagger D W_p & \ W_p^\dagger D W_q  \cr
 W_q^\dagger D W_p & W_q^\dagger D W_q
 \end{pmatrix}
 \equiv \begin{pmatrix}
 \tilde D_{pp} & \tilde D_{pq} \cr
 \tilde D_{qp} & \tilde D_{qq}
 \end{pmatrix} ,
\label{eq:trDt}
\end{equation}
with which the matrix $\bar {\cal U}$ is of the form,
\begin{equation}
 \bar {\cal U}
 = \begin{pmatrix}
 \bar U_{pp}^\dagger \tilde D_{pp} \bar U_{pp}
 +\bar V_{pp}^\dagger \tilde D_{pp}^* \bar V_{pp} &
 \bar U_{pp}^\dagger \tilde D_{pq} \cr
 \tilde D_{qp} \bar U_{pp} & \tilde D_{qq} \end{pmatrix}
 \equiv \begin{pmatrix}
 \bar {\cal U}_{pp} & \bar {\cal U}_{pq} \cr
 \bar {\cal U}_{qp} & \bar {\cal U}_{qq} \end{pmatrix}.
\end{equation}
Since there are no reasons to expect that the transformation matrix
related to the $Q$ space, $\tilde D_{qp}$ or $\tilde D_{qq}$,
is small in any sense,
the number of dimension to calculate the determinant of the norm overlap
in Eq.~(\ref{eq:DUovlp}) cannot be reduced.
However, as it is mentioned in the Appendices in Refs.~\cite{BFH90,YM09},
the model space truncation in terms of $(U,V)$ amplitudes is
possible, which can be naturally derived by the following
treatment in terms of the Thouless amplitude:
We demonstrate it in the Appendix.

On the other hand, if we change the notation and
consider the Thouless form of the state
$|\Phi\rangle$ with respect to the canonical basis $\hat b_k^{}$,
\begin{equation}
 |\Phi\rangle = n \,e^{\hat Z}|\rangle,\quad
 \hat Z\equiv\frac{1}{2}\sum_{k'k}
 Z_{k'k}\hat b_{k'}^\dagger b_k^\dagger,
\end{equation}
with the definition,
\begin{equation}
 Z \equiv(\bar V \bar U^{-1})^*
 = \begin{pmatrix}
 (\bar V_{pp} \bar U_{pp}^{-1})^* & 0 \cr 0 & 0
 \end{pmatrix}  = \begin{pmatrix}
  Z_{pp} & 0 \cr 0 & 0
  \end{pmatrix},
\label{eq:ZUVp}
\end{equation}
the norm overlap~(\ref{eq:DUovlp}) can be easily calculated
(see the next subsection for details) as
\begin{equation}
 \langle \Phi|\hat D|\Phi\rangle
 = e^{i g^0} |\det \bar U|
 \left(\mathstrut \det \left[1+ Z^\dagger Z_D \right]\right)^{1/2},
\label{eq:normZD}
\end{equation}
where the transformed Thouless amplitude $Z_D$ is introduced by
\begin{equation}
 Z_D  \equiv \tilde D Z \tilde D^T \ =\
\begin{pmatrix}
 \tilde D_{pp} Z_{pp} \tilde D_{pp}^T &
 \tilde D_{pp} Z_{pp} \tilde D_{qp}^T \cr
 \tilde D_{qp} Z_{pp} \tilde D_{pp}^T &
 \tilde D_{qp} Z_{pp} \tilde D_{qp}^T
\end{pmatrix}.
\label{eq:trZD}
\end{equation}
The non-trivial point for the model space truncation is
that the amplitude $Z_D$ is not confined within the $P$ space
in contrast to $Z$.
However, the matrix appearing in the norm overlap is of the form,
\begin{equation}
 1+ Z^\dagger Z_D
 = \begin{pmatrix}
 1+ Z_{pp}^\dagger Z_{Dpp} & Z_{pp}^\dagger Z_{Dpq} \cr 0 & 1
 \end{pmatrix},
\label{eq:1+ZZD}
\end{equation}
so that the determinant in Eq.~(\ref{eq:normZD}) can be calculated
within the $P$ space only,
\begin{equation}
  \det\left[ 1+ Z^\dagger Z_D \right]
 =\det\left[ 1+ Z^\dagger_{pp} Z_{Dpp} \right],
\label{eq:ZDtrn}
\end{equation}
where we simply use the notation like $Z_{pp}^\dagger \equiv (Z_{pp})^\dagger$
if there is no confusion.
Namely, the dimension of the determinant is reduced from $M$ to $L_p$,
if one uses the representation in terms of the Thouless amplitude.
In the next subsection we show that not only the norm overlap but also
most part of calculations of the contractions can be done
within the truncated $P$ space for general cases,
and the amount of calculation is greatly reduced
by employing the Thouless amplitudes.

\subsection{Calculation within truncated space}
\label{sec:Caltrunc}

As is discussed in the previous subsections,
the quantity to be calculated is
$\langle \Phi | {\hat O} {\hat D}| \Phi' \rangle$
for an arbitrary operator $\hat O$ with the unitary
transformation ${\hat D}$ of the symmetry operation.
Using the generalized Wick theorem, its evaluation
reduces to calculate the following basic contractions
(or overlaps),
\begin{equation}
\begin{array}{ll}
 \left(\rho^{(c)}_{D}\right)_{l'l} \equiv&
 \langle \Phi|\hat c_{l}^\dagger \hat c_{l'}^{}[\hat D]|\Phi' \rangle,
  \cr
 \left(\kappa^{(c)}_{D}\right)_{l'l} \equiv&
 \langle \Phi|\hat c_{l}^{} \hat c_{l'}^{}[\hat D]|\Phi' \rangle,
  \cr
 \left(\bar\kappa^{(c)}_{D}\right)_{l'l} \equiv&
 \langle \Phi|\hat c_{l}^\dagger \hat c_{l'}^\dagger[\hat D]|\Phi' \rangle,
\label{eq:Dcontc}
\end{array}
\end{equation}
with the definition
\begin{equation}
 [\hat D]\equiv {\hat D}/ \langle \Phi|\hat D|\Phi' \rangle,
\label{eq:defD}
\end{equation}
where the argument $(\bm{x})$ is simply omitted.
In this subsection we develop the efficient method to evaluate
the contractions above as well as the norm overlap
$\langle \Phi|\hat D|\Phi' \rangle$ applying the truncation scheme
explained in the previous subsection.

Thus, we introduce two bases associated with two HFB type states
$|\Phi \rangle$ and $|\Phi' \rangle$,
\begin{equation}
 \hat b^\dagger_k=\sum_l W_{lk} \hat c^\dagger_l,\qquad
 \hat b'^\dagger_k=\sum_l W'_{lk} \hat c^\dagger_l,
\label{eq:twocanob}
\end{equation}
respectively,
with the transformation matrices $W=(W_p,W_q)$ and $W'=(W'_{p'},W'_{q'})$,
where the two bases satisfy
\begin{equation}
 \hat b_k|\Phi\rangle=0,\quad k>L_p \,, \qquad
 \hat b'_{k'}|\Phi'\rangle=0,\quad k'>L_{p'},
\label{eq:btrn}
\end{equation}
namely, the submatrices
$W_p$ and $W'_{p'}$ are $M\times L_p$ and $M\times L_{p'}$, respectively.
The quantity $L_p$ ($L_{p'}$) defines the dimension of the $P$ space
for $|\Phi \rangle$ ($|\Phi' \rangle$).
These bases operators $(\hat b^\dagger,\hat b)$ and $(\hat b'^\dagger,\hat b')$
are practically obtained by diagonalizing the density matrices for
$|\Phi \rangle$ and $|\Phi' \rangle$ like in Eq.~(\ref{eq:Wcanb}),
but one should note that they are not necessarily
the canonical bases if there exist extra degeneracies
for the occupation numbers.
Therefore we call them canonical-like bases.
We introduce the Thouless amplitudes
for these bases $(\hat b^\dagger,\hat b)$ and $(\hat b'^\dagger,\hat b')$
as
\begin{equation}
\begin{array}{lll}
 |\Phi \rangle
 = n\, e^{\hat Z} | \rangle , \quad
 &{\displaystyle \hat Z \ \equiv \ \sum_{k'<k}
 Z_{k'k} \hat b_{k'}^\dagger \hat b_k^\dagger}, \quad
 &Z = -Z^T,  \cr
 |\Phi'\rangle
 = n'\, e^{\hat Z'} | \rangle , \ \ \
 &{\displaystyle \hat Z' \ \equiv \ \sum_{k'<k}
 Z'_{k'k}
    \hat b'^\dagger_{k'} \hat b'^\dagger_k}, \ \ \
 &Z' =\ -Z'^T.
\end{array}
\label{eq:bZZ}
\end{equation}
Here $n$ and $n'$ are normalization constants of
the vacuum states $|\Phi \rangle$ and $|\Phi'\rangle$,
which are not specified here.
Note that the Thouless amplitudes $Z$ ($Z'$) defined
by Eq.~(\ref{eq:bZZ}) can be calculated
from the original $(U,V)$ ($(U',V')$) amplitudes and $W$ ($W'$)
for given state $|\Phi \rangle$ ($|\Phi'\rangle$), and is
essentially $L_p \times L_p$ ($L_{p'} \times L_{p'}$) matrix, i.e.
\begin{equation}
 Z=W^\dagger (VU^{-1})^* W^*=
 \begin{pmatrix}
 Z_{pp} & 0 \cr
 0 & 0 \end{pmatrix},\quad
 Z'=W'^\dagger (V'U'^{-1})^* W'^*=
 \begin{pmatrix}
 Z'_{p'p'} & 0 \cr
 0 & 0 \end{pmatrix}
\end{equation}

The unitary transformation $\hat D$ in Eq.~(\ref{eq:DandG})
induces the transformation between the two bases
$(\hat b^\dagger,\hat b)$ and $(\hat b'^\dagger,\hat b')$,
\begin{equation}
 \hat D \hat b'^\dagger_k \hat D^\dagger
 = \sum_{k'} {\tilde D_{k'k}} \hat b^\dagger_k ,\ \ \
 \hat D^\dagger \hat b_k^\dagger \hat D
 = \sum_{k'}
 {\tilde D_{kk'}^*} \hat b'^\dagger_{k'} ,
\label{eq:Trbb}
\end{equation}
with the definition similarly to Eq.~(\ref{eq:trDt}),
\begin{equation}
 \tilde D \equiv W^\dagger D W' =
 \begin{pmatrix}
 W_p^\dagger D W'_{p'} & W_p^\dagger D W'_{q'} \cr
 W_q^\dagger D W'_{p'} & W_q^\dagger D W'_{q'}
 \end{pmatrix} \equiv
 \begin{pmatrix}
 \tilde D_{pp'} & \tilde D_{pq'} \cr
 \tilde D_{qp'} & \tilde D_{qq'}
 \end{pmatrix},
\label{eq:trD}
\end{equation}
where the transformation matrix $D$ in the original basis
$(\hat c^\dagger,\hat c)$ is defined in Eq.~(\ref{eq:Dgmat}),
and the induced matrix $\tilde D_{p{p'}}$ in the $P$ space, for example,
is now rectangular and a $L_p \times L_{p'}$ matrix.
The action of $\hat D$ on the quasi-particle vacuum $|\Phi'\rangle$
can be calculated as
\begin{equation}
 \hat D |\Phi' \rangle
 = n'\, \exp(\hat D \hat Z' \hat D^\dagger) \hat D |\rangle
 = n'\, e^{\hat Z'_D} |\rangle e^{ig_0} ,
\end{equation}
with
\begin{equation}
 \hat Z'_D \equiv \hat D \hat Z' \hat D^\dagger
 = \sum_{l'<l}
 Z'_{Dl'l} \hat b_{l'}^\dagger \hat b_l^\dagger,
\end{equation}
where the transformed Thouless amplitude
similar to Eq.~(\ref{eq:trZD}) is defined by
\begin{equation}
 Z'_D \equiv \tilde D Z' \tilde D^T =
\begin{pmatrix}
 \tilde D_{pp'} Z'_{p'p'} \tilde D_{pp'}^T
 & \tilde D_{pp'} Z'_{p'p'} \tilde D_{qp'}^T
 \cr
 \tilde D_{qp'} Z'_{p'p'} \tilde D_{pp'}^T
 & \tilde D_{qp'} Z'_{p'p'} \tilde D_{qp'}^T
\end{pmatrix}
\equiv
 \begin{pmatrix}
 Z'_{D{pp}}&Z'_{D{pq}}\cr Z'_{D{qp}}&Z'_{D{qq}}\end{pmatrix},
\label{eq:trZDpq}
\end{equation}
which is not confined in the $P$ space.
Then similarly to Eq.(\ref{eq:ZDtrn})
the norm overlap can be evaluated within the $P$ space as
\begin{eqnarray}
 \langle \Phi |\hat D |\Phi' \rangle
 &=& n^* n'\, e^{ig_0}
  \left( \det \left[1+Z_{pp}^\dagger Z'_{D{pp}}\right]\right)^{1/2} \cr
 &=& n^* n'\,\langle |\hat D|\rangle
  (-1)^{L_p(L_p+1)/2}\,
  {\rm pf} \begin{pmatrix}
  Z'_{D{pp}}&-1\cr 1& Z_{pp}^\dagger\end{pmatrix}.
\label{eq:ovlZDp}
\end{eqnarray}
Namely the dimension of matrix is reduced from $M$ to $L_p$.

The basic contractions can be calculated
through the canonical-like basis $(\hat b^\dagger,\hat b)$,
\begin{equation}
 \rho^{(c)}_{D} = W \rho^{(b)}_{D} W^\dagger,\qquad
 \kappa^{(c)}_{D} = W \kappa^{(b)}_{D} W^T,\qquad
 \bar\kappa^{(c)}_{D} = W^* \bar\kappa^{(b)}_{D} W^\dagger,
\label{eq:cbDcont}
\end{equation}
where
\begin{equation}
\begin{array}{ll}
 \left(\rho^{(b)}_{D}\right)_{k'k} \equiv&
 \langle \Phi|\hat b_{k}^\dagger \hat b_{k'}^{}[\hat D]|\Phi' \rangle
 =\left( Z'_D \left[ 1+Z^\dagger Z'_D\right]^{-1} Z^\dagger
  \right)_{k'k},
\cr
 \left(\kappa^{(b)}_{D}\right)_{k'k} \equiv&
 \langle \Phi|\hat b_{k}^{} \hat b_{k'}^{}[\hat D]|\Phi' \rangle
 =\left( Z'_D \left[ 1+Z^\dagger Z'_D\right]^{-1} \right)_{k'k},
\cr
 \left(\bar\kappa^{(b)}_{D}\right)_{k'k} \equiv&
 \langle \Phi|\hat b_{k}^\dagger \hat b_{k'}^\dagger[\hat D]|\Phi' \rangle
 =\left(\left[ 1+Z^\dagger Z'_D\right]^{-1} Z^\dagger \right)_{k'k}.
\label{eq:bDcont}
\end{array}
\end{equation}
Using the corresponding equation to~(\ref{eq:1+ZZD}),
\begin{equation}
 \bar\kappa^{(b)}_{D}
 = \begin{pmatrix}
 [1+ Z_{pp}^\dagger Z_{Dpp}]^{-1} Z_{pp}^\dagger & 0 \cr 0 & 0 \end{pmatrix}
 \equiv \begin{pmatrix}
  \bar\kappa^{(b)}_{Dpp} & 0 \cr 0 & 0 \end{pmatrix},
\label{eq:bkappapp}
\end{equation}
which has only the $P$ space components.
Further using the identities
\begin{equation}
\rho^{(b)}=Z'_D \bar\kappa^{(b)}_D,\qquad
\kappa^{(b)}_D=Z'_D - \rho^{(b)}_D Z'_D=Z'_D-Z'_D\bar\kappa^{(b)}_D Z'_D,
\label{eq:rhokapbybkap}
\end{equation}
which can be easily confirmed by Eq.~(\ref{eq:bDcont}), we have
\begin{equation}
\begin{array}{ll}
 \rho^{(c)}_{D} &= D W'_{p'}\,(Z'_{p'p'} \tilde D_{pp'}^T
 \bar\kappa^{(b)}_{Dpp})\,W_p^\dagger,
\cr
 \kappa^{(c)}_{D} &= D W'_{p'}\,
 (Z'_{p'p'} -Z'_{p'p'}\tilde D_{pp'}^T
 \bar\kappa^{(b)}_{Dpp} \tilde D_{pp'} Z'_{p'p'})\,W'^T_{p'}D^T,
\cr
 \bar\kappa^{(c)}_{D} &= W^*_p\,(\bar\kappa^{(b)}_{Dpp})\,W^\dagger_{p}.
\end{array}
\label{eq:rhokappp}
\end{equation}
Namely, most of the calculations, i.e.,
the part in parentheses in Eq.~(\ref{eq:rhokappp}),
can be done within the $P$ space.

In order to make reduced calculations within the $P$ space
more systematically and to enable a generalization,
which is discussed in the next subsection,
we use the following property of the basis truncation
defined in Eq.~(\ref{eq:btrn});
\begin{equation}
 \hat b_k \hat D|\Phi'\rangle
 = \hat D \sum_{k'}
 {\tilde D_{kk'}} \hat b'_{k'} |\Phi' \rangle
 = \sum_{k_{p'}=1}^{L_{p'}} {(\tilde \tau_{Dp'}^{})_{kk_{p'}}} \hat b_{k_{p'}}
 \hat D |\Phi' \rangle ,
\label{eq:bConTrick}
\end{equation}
where a new $M \times L_{p'}$ matrix $\tilde \tau_{Dp'}^{}$ is defined by
\begin{equation}
 (\tilde \tau_{Dp'}^{})_{kk_{p'}} \equiv
 \sum_{k'_{p'}=1}^{L_{p'}}
 \tilde D_{kk'_{p'}}
 \left(\tilde D_{P'}^{-1}\right)_{k'_{p'} k_{p'}^{}},
 \quad \mbox{i.e.},\quad
 \tilde \tau_{Dp'}^{}=W^\dagger DW'_{p'}\tilde D_{P'}^{-1} .
\label{eq:tauD}
\end{equation}
Here we have introduced an auxiliary $L_{p'} \times L_{p'}$ square submatrix
$\tilde D_{P'}$ of $\tilde D$, and its inverse $\tilde D_{P'}^{-1}$, i.e.,
\begin{equation}
 \tilde D_{P'}=(\tilde D_{kk'};\,k,k'=1,2,...,L_{p'}),
\end{equation}
which should not be confused with the $L_p\times L_{p'}$ submatrix
$\tilde D_{pp'}$ in Eq.~(\ref{eq:trD})
(of course, $\tilde D_{pp'}$ and $\tilde D_{P'}$ coincide if $L_p=L_{p'}$).
The $P$ space should be chosen in such a way that the matrix
$\tilde D_{P'}$ has its inverse.
From our experiences this requirement is usually satisfied
without any special treatments as long as the transformation includes
the rotation as in the case of the angular momentum projection.
While a problem may occurs if the two wave functions
$|\Phi\rangle$ and $|\Phi'\rangle$ have different symmetries,
the rotation strongly mixes them and the rank of
matrix $\tilde D_{P'}$ does not usually reduce.
By using the property in Eq.~(\ref{eq:bConTrick}),
the contractions for the original basis can be calculated as follows;
\begin{equation}
\begin{array}{lll}
 \rho^{(c)}_{D} =& \tau_{Dp'}^{} \,\rho^{(b)}_{Dp'p}\, \eta_p^\dagger
 &=DW'_{p'}\tilde D_{P'}^{-1} \,\rho^{(b)}_{Dp'p}\, W_p^\dagger,
\cr
 \kappa^{(c)}_{D} =& \tau_{Dp'}^{} \,\kappa^{(b)}_{Dp'p'}\, \tau_{Dp'}^T
 &=DW'_{p'}\tilde D_{P'}^{-1} \,\kappa^{(b)}_{Dp'p'}\, D_{P'}^{-T} W'^T_{p'}D^T,
\cr
 \bar\kappa^{(c)}_{D} =& \eta_p^* \,\bar\kappa^{(b)}_{Dpp}\, \eta_p^\dagger
 &=W_p^* \, \bar\kappa^{(b)}_{Dpp}\, W_p^\dagger,
\label{eq:cbDcontrn}
\end{array}
\end{equation}
where a $M\times L_{p'}$ matrix $\tau_{Dp'}^{}$ and
a $M\times L_{p}$ matrix $\eta_{p}$ are defined by
\begin{equation}
 \tau_{Dp'}^{}\equiv W \tilde \tau_{Dp'}
 =DW'_{p'}\tilde D_{P'}^{-1},\quad
 \eta_p \equiv W_p.
\label{eq:tauDtrn}
\end{equation}
The reduced contractions for the $(\hat b^\dagger,\hat b)$ basis
in Eq.~(\ref{eq:cbDcontrn}) are nothing else but their
$L_{p'}\times L_p$, $L_{p'}\times L_{p'}$, and $L_p\times L_p$ submatrices,
respectively;
\begin{equation}
\begin{array}{ll}
 \rho^{(b)}_{Dp'p} &\equiv
 \left((\rho^{(b)}_D)_{kk'};\,k=1,2,...,L_{p'},k'=1,2,...,L_p \right),
\cr
 \kappa^{(b)}_{Dp'p'} &\equiv
 \left((\kappa^{(b)}_D)_{kk'};\,k,k'=1,2,...,L_{p'} \right),
\cr
 \bar\kappa^{(b)}_{Dpp} &\equiv
 \left((\bar\kappa^{(b)}_D)_{kk'};\,k,k'=1,2,...,L_{p} \right),
\label{eq:bDcontrn}
\end{array}
\end{equation}
which can be evaluated within the $P$ space.
This is because they are more explicitly written as,
\begin{equation}
\begin{array}{ll}
 \bar\kappa^{(b)}_{Dpp} &=
 \left[ 1+Z_{pp}^\dagger Z'_{Dpp}\right]^{-1} Z_{pp}^\dagger,
\quad
 \rho^{(b)}_{Dp'p} = Z'_{Dp'p} \bar\kappa^{(b)}_{Dpp},\quad
 \kappa^{(b)}_{Dp'p'} =
 Z'_{Dp'p'} - Z'_{Dp'p} \bar\kappa^{(b)}_{Dpp} Z'_{Dpp'},
\end{array}
\label{eq:bDcontrnx}
\end{equation}
where the subblock matrices of $Z'_D$ are defined by
\begin{equation}
\begin{array}{l}
 Z'_{Dp'p'} \equiv
 \left((Z'_D)_{kk'};k,k'=1,2,...,L_{p'}\right)
 =\tilde D_{P'}Z'_{p'p'}\tilde D_{P'}^T,\cr
 Z'_{Dp'p} \equiv
 \left((Z'_D)_{kk'};k=1,2,...,L_{p'},k'=1,2,...,L_p\right)
 =\tilde D_{P'}Z'_{p'p'}\tilde D_{pp'}^T,\cr
 Z'_{Dpp'} \equiv
 \left((Z'_D)_{kk'};k=1,2,...,L_p,k'=1,2,...,L_{p'}\right)
 =\tilde D_{pp'}Z'_{p'p'} \tilde D_{P'}^T.
\end{array}
\label{eq:ZDsubp}
\end{equation}
It is now clear that the matrix $\tilde D_{P'}$ and its inverse appearing
in the matrix $\tau_{Dp'}^{}$ in Eq.~(\ref{eq:tauDtrn}) are auxiliary
and introduced just for the sake of convenience of calculation.
In fact it is confirmed by
Eqs.~(\ref{eq:cbDcontrn}), (\ref{eq:bDcontrnx}), and~(\ref{eq:ZDsubp})
that the basic contractions for the original basis $(\hat c^\dagger,\hat c)$
are independent of them.

With these basic contractions for the $(\hat b^\dagger,\hat b)$ basis,
overlaps of arbitrary one-body operators can be easily calculated.
For the particle-hope (p-h) type operator, $\hat F$,
and particle-particle (p-p) or hole-hole (h-h) type operator,
$\hat G^\dagger$ or $\hat G$,
\begin{equation}
 \hat F = \sum_{l_1 l_2} F_{l_1 l_2}
   \hat c_{l_1}^\dagger \hat c_{l_2} ,\qquad
 \hat G^\dagger = \frac{1}{2}
  \sum_{l_1 l_2} G_{l_1 l_2}
   \hat c_{l_1}^\dagger \hat c_{l_2}^\dagger ,
\end{equation}
with antisymmetric matrix elements $G^T  = -G$,
\begin{equation}
\begin{array}{lll}
 \langle \Phi|\hat F [\hat D]| \Phi' \rangle
 &=\ {\rm Tr} \{\rho^{(c)}_D  F\}
 &=\ {\rm Tr} \{\rho^{(b)}_{Dp'p }  F_D^{pp'} \}, \vspace*{1mm} \cr
 \langle \Phi |\hat G [\hat D]| \Phi' \rangle
 &=\ {\displaystyle \frac{1}{2}} {\rm Tr} \{ \kappa^{(c)}_D G^\dagger\}
 &=\ {\displaystyle \frac{1}{2}}
 {\rm Tr} \{ \kappa^{(b)}_{Dp'p'} \bar G_D^{p'p'} \}, \vspace*{1mm} \cr
 \langle \Phi|\hat G^\dagger [\hat D]| \Phi' \rangle
 &=\ {\displaystyle \frac{1}{2}} {\rm Tr} \{\bar\kappa^{(c)}_D G\}
 &=\ {\displaystyle \frac{1}{2}}
 {\rm Tr} \{\bar\kappa^{(b)}_{Dpp } G^{pp} \},
\label{eq:FGbtrn}
\end{array}
\end{equation}
where the $P$ space matrix elements for $\hat F$, $\hat G^\dagger$ and $\hat G$
are defined by using the quantities in Eq.~(\ref{eq:tauDtrn}),
\begin{equation}
\begin{array}{lll}
 F_D^{pp'} &\equiv \ \eta_p^\dagger F \tau_{Dp'}^{}
 &=\ W_p^\dagger F DW'_{p'} \tilde D_{P'}^{-1}, \vspace*{1mm} \cr
 \bar G_D^{p'p'} &\equiv \ \tau_{Dp'}^T G^\dagger \tau_{Dp'}^{}
 &=\ \tilde D_{P'}^{-T}W'^T_{p'}D^T G^\dagger DW'_{p'} \tilde D_{P'}^{-1},
 \vspace*{1mm} \cr
 G^{pp} &\equiv \ \eta_p^\dagger G \eta_p^*
 &=\ W_p^\dagger G W_p^* .
\label{eq:FGtp}
\end{array}
\end{equation}
In the actual applications of the angular momentum projection,
the operator is a spherical tensor,
e.g., $\hat G^\dagger=\hat G^\dagger_{\lambda\mu}$,
and its matrix elements in the original basis satisfy
\begin{equation}
 D^T(\omega) G_{\lambda \mu}^\dagger D(\omega) =
 \sum_{\mu'} D^\lambda_{\mu \mu'}(\omega)\, G_{\lambda \mu'}^\dagger,
\end{equation}
where $D^\lambda_{\mu \mu'}(\omega)$ is the Wigner $D$-function,
and then
\begin{equation}
 (\bar G_{\lambda \mu})_D^{p'p'}
 = \sum_{\mu'} D^\lambda_{\mu \mu'}(\omega)\,
 \tilde D_{P'}^{-T}
 (G_{\lambda \mu'}^{p'p'})^\dagger
 \tilde D_{P'}^{-1} , \quad
 G_{\lambda \mu}^{p'p'} \equiv W'^\dagger_{p'}
 G_{\lambda \mu} W'^*_{p'},
\end{equation}
which can be calculated within the $P$ space.
The task is to evaluate the overlap at each integration
mesh point in the parameter space, which requires
$O(M^3)$ operations (matrix multiplications) for one-body operators
in the original basis.  Now it reduces to
$O(ML_p^2)$ for the p-h type operator $\hat F$
and $O(L_p^3)$ for the p-p or h-h operator $\hat G^\dagger$ or $\hat G$
in the truncation scheme ($L_p \sim L_{p'}$).

In this paper, we employ separable type schematic interactions.
By using the generalized Wick Theorem,
we have, for the p-h type interaction,
\begin{eqnarray}
 \langle \Phi | :\hat F_1 \hat F_2 :[\hat D] |\Phi'\rangle
 &=& {\rm Tr} \{\rho^{(c)}_D  F_1\} {\rm Tr} \{ \rho^{(c)}_D F_2 \}
 -{\rm Tr} \{\rho^{(c)}_D  F_1 \rho^{(c)}_D F_2 \}
 +{\rm Tr} \{\bar\kappa^{(c)}_D F_1 \kappa^{(c)}_D  F_2^T \}
\nonumber\\
 &=& {\rm Tr} \{\rho^{(b)}_{Dp'p}  F_{1D}^{pp'} \}
  {\rm Tr} \{ \rho^{(b)}_{Dp'p} F_{2D}^{pp'} \}
\nonumber \\
 &&-{\rm Tr} \{\rho^{(b)}_{Dp'p}  F_{1D}^{pp'}
 \rho^{(b)}_{Dp'p} F_{2D}^{pp'} \} +{\rm Tr} \{ \bar\kappa^{(b)}_{Dpp}
 F_{1D}^{pp'} \kappa^{(b)}_{Dp'p'} F_{2D}^{pp' T} \},
\label{eq:FFtrn}
\end{eqnarray}
where $:\ :$ denotes the normal ordering,
and for the p-p or h-h type interaction,
\begin{eqnarray}
 \langle \Phi |\hat G_1^\dagger \hat G_2 [\hat D] |\Phi'\rangle
 &=& \frac{1}{4} \left[
 {\rm Tr} \{\bar\kappa^{(c)}_D G_1\} {\rm Tr} \{ \kappa^{(c)}_D G_2^\dagger\}
 +2{\rm Tr} \{ \rho^{(c)}_D G_1 \rho^{(c)T}_D G_2^\dagger \}
 \right]
\nonumber\\
 &=&  \frac{1}{2}{\rm Tr} \{\bar\kappa^{(b)}_{Dpp} G_1^{pp} \}
  \frac{1}{2}{\rm Tr} \{ \kappa^{(b)}_{Dp'p'} \bar G_{2D}^{p'p'} \}
 +\frac{1}{2}{\rm Tr} \{ \rho^{(b)}_{Dp'p} G_1^{pp}
  \rho^{(b)T}_{Dp'p} \bar G_{2D}^{p'p'} \}.\quad
\label{eq:GGtrn}
\end{eqnarray}
Thus, the basic number of operations to calculate the overlap
of the separable type interactions is
essentially the same as those of one-body operators,
and can be evaluated much faster than the generic two-body
interaction (as long as the number of
the separable force components are not so large).

For the generic two-body interaction,
there are four single-particle indices with two density matrices $\rho$
or with two pairing tensors $\kappa$ and $\bar\kappa$.
As is shown in Eq.~(\ref{eq:cbDcontrn}), the two among the four
indices are accompanied with the rotation matrix $D$,
and therefore the reduction of the number of operations
from $O(M^4)$ to $O(M^2L_p^2)$ is expected.

\subsection{Truncation with respect to particle-hole vacuum}
\label{sec:truncphv}

As it is demonstrated in the previous subsection,
the use of the Thouless amplitude
with respect to the nucleon vacuum, Eq.~(\ref{eq:bZZ}),
allows us to dramatically reduce
the number of dimension of matrices in the calculation.
However, the problem occurs if one takes a limit of
vanishing pairing correlations.
This is because the amplitude $U\rightarrow 0$
for the hole (occupied) orbits in the limit,
and then the Thouless amplitude $Z$ diverges.
Moreover, the Thouless form in Eq.~(\ref{eq:ThoulZ})
can be applied only for the case where the HFB type states
$|\Phi\rangle$ and $|\Phi'\rangle$ are not orthogonal to the nucleon
vacuum, i.e., for the ground states of even-even nuclei.
In order to avoid these problems and to generalize the formulation,
we introduce the Thouless amplitude
with respect to the p-h vacuum (Slater determinant)
in place of the nucleon vacuum.
Although this makes the formulation more complicated,
we have an additional merit;
the contribution of core composed of the fully occupied orbits,
whose occupation probability is almost one, can be separated
and the amount of calculation is further reduced.
This effect is considerable especially for heavy nuclei.

Thus, for the two HFB type states $|\Phi\rangle$ and $|\Phi'\rangle$,
we introduce the particle-hole vacuums (Slater determinants),
which are composed of $N$ canonical-like basis orbits
with highest occupation probabilities,
\begin{equation}
 |\phi\rangle = \prod_{k=1}^N \hat b^\dagger_k \,|\rangle,\qquad
 |\phi'\rangle = \prod_{k=1}^N \hat b'^\dagger_k \,|\rangle,
\label{eq:phvac}
\end{equation}
where $N$ is the particle (neutron or proton) number.
Note that the index of the canonical-like bases,
$(\hat b^\dagger_i,\hat b^{}_i)$ and $(\hat b'^\dagger_i,\hat b'^{}_i)$,
introduced in the previous subsections,
Eqs.~(\ref{eq:canb}) and~(\ref{eq:Wcanb}), is in descending order of
the occupation probabilities.  Therefore,
$|\Phi\rangle \rightarrow |\phi\rangle$
and $|\Phi'\rangle \rightarrow |\phi'\rangle$
in the limit of vanishing pairing correlations,
if the two states $|\Phi\rangle$ and $|\Phi'\rangle$ are normalized
and their phases are suitably chosen.
More precisely, when there exists an unbroken symmetry, e.g., the parity,
the $N$ hole orbits should be chosen
so that the states $|\Phi\rangle$ and $|\phi\rangle$
($|\Phi'\rangle$ and $|\phi'\rangle$)
belong to the same symmetry representation.
Corresponding to the p-h vacuums in Eq.~(\ref{eq:phvac}),
the canonical particle-hole operators $(\hat a^\dagger,\hat a)$,
which satisfy
\begin{equation}
 \hat a_k |\phi\rangle=0 \quad (k=1,2,...,M),\qquad
 \hat a'_k |\phi'\rangle=0 \quad (k=1,2,...,M),
\end{equation}
are defined by
\begin{equation}
 a^\dagger_k= \left\{\begin{array}{ll}
 b^{}_k \quad& (1\le k \le N)\cr
 b^\dagger_k \quad& (N+1 \le k \le M)\end{array}\right.,\qquad
 a'^\dagger_k= \left\{\begin{array}{ll}
 b'^{}_k \quad& (1\le k \le N)\cr
 b'^\dagger_k \quad& (N+1 \le k \le M)\end{array}\right. .
\label{eq:phopt}
\end{equation}
The relations between these particle-hole bases
and the original basis $(\hat c^\dagger,\hat c)$ are given by
general Bogoliubov transformations,
\begin{equation}
 \hat a_k^\dagger
 = \sum_{l} \left[ (u_a)_{lk} \hat c_{l}^\dagger
 + (v_a)_{lk} \hat c_{l} \right] , \ \ \
 \hat a'^\dagger_k
 = \sum_{l} \left[ (u'_a) {lk} \hat c_{l}^\dagger
 + (v'_a)_{lk} \hat c_{l} \right] ,
\end{equation}
where the Bogoliubov amplitudes $(u_a, v_a)$
and $(u'_a, v'_a)$ are simply given by $W$ and $W'$ matrices
but specified by the following particle-hole block structure,
\begin{equation}
 \left\{\begin{array}{l}
 u_a = \begin{pmatrix} 0 & W_m \end{pmatrix} \cr
 v_a = \begin{pmatrix} W_i^* & 0 \end{pmatrix}
 \end{array} \right. ,\qquad
 \left\{\begin{array}{l}
 u'_a= \begin{pmatrix} 0 & W'_{m'} \end{pmatrix} \cr
 v'_a= \begin{pmatrix} W'^{*}_{i'} & 0 \end{pmatrix}
 \end{array}\right. ,\qquad
\end{equation}
where $W_i$ and $W'_{i'}$ are the hole part of matrices and of $M\times N$,
while $W_m$ and $W'_{m'}$ are the particle part of matrices
and of $M\times (M-N)$.
This particle-hole decomposition should not be confused with
the $P$ and $Q$ space decomposition in Eq.~(\ref{eq:WUVpq}),
and inequalities $N \le L_p$ and $N \le L_{p'}$ should be satisfied.

Now we assume that the HFB type states are normalized,
and define their Thouless forms with respect to the p-h vacuums.
In this subsection we change the notation,
and use $Z$ for the Thouless amplitudes for this representation:
\begin{equation}
 |\Phi \rangle = n\, \exp\Bigl[
  \sum_{k<k'} Z_{kk'} \hat a_k^\dagger \hat a_{k'}^\dagger
 \Bigr] | \phi \rangle, \qquad
 |\Phi' \rangle = n'\, \exp\Bigl[
  \sum_{k<k'} Z'_{kk'} \hat a'^\dagger_k \hat a'^\dagger_{k'}
 \Bigr] | \phi'\rangle.
\label{eq:Thph}
\end{equation}
The Thouless amplitudes and the normalization constants
in this representation are calculated by
\begin{equation}
Z=(V_aU_a^{-1})^*,\qquad
Z'=(V'_aU'^{-1}_a)^*,
\end{equation}
\begin{equation}
 n=e^{i\theta}\left(\det{U_a^*}\right)^{1/2},\qquad
 n'=e^{i\theta'}\left(\det{U'^*_a}\right)^{1/2},
\label{eq:Zanor}
\end{equation}
through the Bogoliubov amplitudes $(U_a,V_a)$ between
the quasiparticle basis $(\hat \beta^\dagger,\hat \beta)$ and
the p-h basis $(\hat a^\dagger,\hat a)$,
\begin{equation}
 \hat \beta^\dagger_k
 = \sum_{k'} \left[
  (U_a)_{k'k} \hat a_{k'}^\dagger +(V_a)_{k'k} \hat a_{k'}^{}
 \right],
 \qquad
 \hat \beta'^\dagger_k
 = \sum_{k'} \left[
  (U'_a)_{k'k} \hat a'^\dagger_{k'} +(V'_a)_{k'k} \hat a'_{k'}
 \right],
\label{eq:trUaVa}
\end{equation}
and they are written as
\begin{equation}
 U_a=\begin{pmatrix} W_i^T V \cr W_m^\dagger U \end{pmatrix},\quad
 V_a=\begin{pmatrix} W_i^\dagger U \cr W_m^T V \end{pmatrix},\quad
 U'_a=\begin{pmatrix} W'^T_{i'} V' \cr W'^\dagger_{m'}U' \end{pmatrix},\quad
 V'_a=\begin{pmatrix} W'^\dagger_{i'} U' \cr W'^T_{m '}V' \end{pmatrix},
\label{eq:UVa}
\end{equation}
where $(U,V)$ and $(U',V')$ are the Bogoliubov amplitudes with respect to
the original basis $(\hat c^\dagger, \hat c)$
for $|\Phi\rangle$ and $|\Phi'\rangle$, respectively.
As it clear from Eq.~(\ref{eq:UVa}),
$U_aU_a^\dagger \rightarrow 1$, $V_a^*V_a^T\rightarrow 0$
for all orbits in the limit of no pairing correlations,
and then the Thouless amplitude in this representation does not diverge
but vanishes, $Z\rightarrow 0$.
The same is true for $Z'$ and $|\Phi'\rangle$.

The transformation between the two p-h bases $(\hat a^\dagger,\hat a)$
and $(\hat a'^\dagger,\hat a')$ induced by the symmetry operation $\hat D$
is also given by a general Bogoliubov transformation,
\begin{equation}
 \hat D \hat a'^\dagger_{k'} \hat D^\dagger
 = \sum_k \left[ (X_D)_{kk'} \hat a_k^\dagger +(Y_D)_{kk'} \hat a_k^{} \right],
\label{eq:trDaa}
\end{equation}
with the amplitudes defined by
\begin{equation}
\begin{array}{ll}
 X_D \equiv u_a^\dagger D u'_{a'} + v_a^\dagger D^* v'_{a'}
 =& \begin{pmatrix}
 \tilde D^*_{i i'} & 0 \cr 0 & \tilde D_{m m'}
 \end{pmatrix},
 \vspace*{1mm}\cr
 Y_D \equiv v_a^T D u'_{a'} + u_a^T D^* v'_{a'}
 =& \begin{pmatrix}
 0 & \tilde D_{im'} \cr \tilde D^*_{mi'} & 0
 \end{pmatrix},
\end{array}
\label{eq:XYD}
\end{equation}
where the matrix $\tilde D$ is the same as that in Eq.~(\ref{eq:trD})
but divided into the p-h block form,
\begin{equation}
 \tilde D \equiv  W^\dagger D W'
 = \begin{pmatrix}
 W_i^\dagger D W'_{i'} & W_i^\dagger D W'_{m'} \cr
 W_m^\dagger D W'_{i'} & W_m^\dagger D W'_{m'}
 \end{pmatrix} \equiv
 \begin{pmatrix}
 \tilde D_{ii'} & \tilde D_{im'} \cr
 \tilde D_{mi'} & \tilde D_{mm'}
 \end{pmatrix}.
\label{eq:trDph}
\end{equation}
Combining Eqs.~(\ref{eq:trUaVa}) and~(\ref{eq:trDaa}),
the transformed quasiparticle operator for the state $\hat D|\Phi'\rangle$
is expressed as
\begin{equation}
 \hat D \hat \beta'^\dagger_k \hat D^\dagger
 = \sum_{k'} \left[
  (U_{aD}')_{k'k} \hat a^\dagger_{k'} +(V_{aD}')_{k'k} \hat a_{k'}
 \right],
\label{eq:trUaVaD}
\end{equation}
with
\begin{equation}
\begin{array}{l}
 {U_{aD}'}=X_D^{}U'_a+Y_D^* V'_a
 =\left[ X_D^* +Y_D^{}Z' \right]^* U'_a ,\cr
 {V_{aD}'}=X_D^* V'_a+Y_D^{}U'_a
 =\left[ X_D Z'+Y_D^* \right]^* U'_a , \end{array}
\label{eq:UaVaD}
\end{equation}
from which the Thouless form of the transformed state is obtained;
\begin{equation}
 \hat D|\Phi'\rangle
 =n'\,e^{i\Theta(\hat D)}
  \left(\det (U'_{aD}U'^{-1}_a)^*\right)^{1/2}\,
  \exp\left[\sum_{k<k'} (Z'_D)_{kk'}
  \hat a_k^\dagger \hat a_{k'}^\dagger
 \right]|\phi\rangle ,
\label{eq:trDPhip}
\end{equation}
where the phase $\Theta(\hat D)$ coming from the transformation
is introduced in Eqs.~(\ref{eq:Dphase}) and~(\ref{eq:Dmean}),
and the new Thouless amplitude $Z'_D$ is defined by
\begin{equation}
 Z'_D \equiv \left(V'_{aD} U'^{-1}_{aD} \right)^*
 =\left[X_D^{}Z'+Y_D^* \right] \left[X_D^* +Y_D^{}Z' \right]^{-1}.
\label{eq:ZDpXY}
\end{equation}

Introducing two new antisymmetric matrices,
\begin{equation}
 S_D^\dagger \equiv X_D^{-*}Y_D = -S_D^*,\qquad
 {\tilde S}_D \equiv \left(Y_D X_D^{-1}\right)^*=-{\tilde S}_D^T,
\label{eq:defSD}
\end{equation}
the Thouless amplitude of the transformed state
in Eq.~(\ref{eq:ZDpXY}) can be written as
\begin{equation}
 Z'_D=X_D^{-\dagger}
 Z'\left[1+S_D^\dagger Z'\right]^{-1} X_D^{-*}+{\tilde S}_D,
\label{eq:ZDpmod}
\end{equation}
and the norm overlap is calculated as
\begin{eqnarray}
 \langle \Phi| \hat D |\Phi' \rangle
 &=& n^* n'\, e^{i\Theta(\hat D)}
 \left(\det \left[X_D^*+Y_D^{}Z' \right]\right)^{1/2}
 \left(\det \left[ 1+Z^\dagger {Z'_D} \right]\right)^{1/2} \cr
 &=& n^* n'\, e^{i\Theta(\hat D)} \left(\det X_D^* \right)^{1/2}
 \Bigl(\det \left[1+S_D^\dagger Z' \right]_{p'p'}\Bigr)^{1/2}
 \left(\det \left[1+Z^\dagger {Z'_D} \right]_{pp}\right)^{1/2} \cr
 &=& n^* n'\, \langle\phi|\hat D|\phi'\rangle \,
 (-1)^{L_{p'}(L_{p'}+1)/2}\, (-1)^{L_p(L_p+1)/2} \cr
 &&\qquad\times\,
  {\rm pf} \begin{pmatrix}
  Z'_{p'p'}&-1\cr 1& S_{Dp'p'}^\dagger\end{pmatrix}
  {\rm pf} \begin{pmatrix}
  Z'_{D{pp}}&-1\cr 1& Z_{pp}^\dagger\end{pmatrix},
\label{eq:DPhiovlp}
\end{eqnarray}
where the following identity for the norm overlap for the p-h vacuums
is used;
\begin{equation}
 e^{i\Theta(\hat D)} \left(\det X_D^* \right)^{1/2}
 =\langle \phi| \hat D |\phi' \rangle.
\label{eq:DovlpSD}
\end{equation}
Taking into account the fact that
\begin{equation}
 Z'_{Dpp}=(X_D^{-\dagger})_{pp'}
 Z'_{p'p'}\left[1+S_{Dp'p'}^\dagger Z'_{p'p'}\right]^{-1}
 (X_D^{-*})_{p'p}+{\tilde S}_{Dpp},
\label{eq:ZDpmodpp}
\end{equation}
the norm overlap in Eq.~(\ref{eq:DPhiovlp}) can be calculated
within the $P$ space,
if the quantities $(X_D^{-1})_{pp'}$, $S_{Dp'p'}^\dagger$,
${\tilde S}_{Dpp}$, and $\langle \phi| \hat D |\phi' \rangle$
can be calculated easily.  This is actually the case,
because their explicit forms can be written as
\begin{equation}
 X_D^{-1} =
\begin{pmatrix}
  \tilde D_{ii'}^{-*} & 0 \cr
  0 & \tilde D_{mm'}^{-1}
\end{pmatrix}=
\begin{pmatrix}
  \tilde D_{ii'}^{-*} & 0 \cr
  0 & \tilde D_{mm'}^\dagger-\tilde D_{im'}^\dagger
  \tilde D_{ii'}^{-\dagger} {\tilde D_{mi'}}^\dagger
\end{pmatrix},
\end{equation}
\begin{equation}
 S_D^\dagger =
\begin{pmatrix}
  0 & \tilde D_{ii'}^{-1} \tilde D_{im'} \cr
  -\tilde D_{im'}^T \tilde D_{ii'}^{-T} & 0
\end{pmatrix}, \qquad
 {\tilde S}_D =
\begin{pmatrix}
  0 & -\tilde D_{ii'}^{-T} {\tilde D_{mi'}^T} \cr
   \tilde D_{mi'} \tilde D_{ii'}^{-1} & 0
\end{pmatrix},
\end{equation}
and
\begin{equation}
 \langle \phi| \hat D |\phi' \rangle
 =\det{\tilde D_{ii'}},
\label{eq:DovlpSDh}
\end{equation}
so that the matrix manipulations are confined in the hole space,
which is smaller than (or equal to) the $P$ space.

As for the contractions, those for the p-h basis
$(\hat a^\dagger,\hat a)$ can be calculated
in terms of the new Thouless amplitudes introduced in this subsection,
$Z$ and $Z'_D$ in Eqs.~(\ref{eq:Thph}) and~(\ref{eq:trDPhip}),
as
\begin{equation}
\begin{array}{ll}
 \left(\rho^{(a)}_{D}\right)_{k'k} \equiv&
 \langle \Phi|\hat a_{k}^\dagger \hat a_{k'}^{}[\hat D]|\Phi' \rangle
 =\left( Z'_D \left[ 1+Z^\dagger Z'_D\right]^{-1} Z^\dagger
  \right)_{k'k},
\cr
 \left(\kappa^{(a)}_{D}\right)_{k'k} \equiv&
 \langle \Phi|\hat a_{k}^{} \hat a_{k'}^{}[\hat D]|\Phi' \rangle
 =\left( Z'_D \left[ 1+Z^\dagger Z'_D\right]^{-1} \right)_{k'k},
\cr
 \left(\bar\kappa^{(a)}_{D}\right)_{k'k} \equiv&
 \langle \Phi|\hat a_{k}^\dagger \hat a_{k'}^\dagger[\hat D]|\Phi' \rangle
 =\left(\left[ 1+Z^\dagger Z'_D\right]^{-1} Z^\dagger \right)_{k'k}.
\label{eq:aDcont}
\end{array}
\end{equation}
Their structures in terms of $Z$ and $Z'_D$ matrices are the same
as those for the $(\hat b^\dagger,\hat b)$ basis in the previous subsection.
Namely, $\bar\kappa^{(a)}_{D}$ has the same block form
as in Eq.~(\ref{eq:bkappapp}), and the same identities
as in Eq.~(\ref{eq:rhokapbybkap}) hold.
Therefore, their reduced contractions,
\begin{equation}
\begin{array}{ll}
 \rho^{(a)}_{Dp'p} &\equiv
 \left((\rho^{(a)}_D)_{kk'};\,k=1,2,...,L_{p'},k'=1,2,...,L_p \right),
\cr
 \kappa^{(a)}_{Dp'p'} &\equiv
 \left((\kappa^{(a)}_D)_{kk'};\,k,k'=1,2,...,L_{p'} \right),
\cr
 \bar\kappa^{(a)}_{Dpp} &\equiv
 \left((\bar\kappa^{(a)}_D)_{kk'};\,k,k'=1,2,...,L_{p} \right),
\label{eq:aDcontrn}
\end{array}
\end{equation}
can be evaluated within the $P$ space.
By using the definition in Eq.~(\ref{eq:phopt}),
the contractions for the $(\hat b^\dagger,\hat b)$ basis
are related to those for the $(\hat a^\dagger,\hat a)$ basis;
\begin{equation}
 \rho^{(b)}_D=
\begin{pmatrix}
 1_{ii}- {\rho^{(a)T}_{Dii}} & \bar\kappa^{(a)}_{Dim} \cr
  \kappa^{(a)}_{Dmi} & \rho^{(a)}_{Dmm}
\end{pmatrix},\
 \kappa^{(b)}_D=
\begin{pmatrix}
  \bar\kappa^{(a)}_{Dii} & - {\rho^{(a)T}_{Dmi}} \cr
  {\rho^{(a)}_{Dmi}} & \kappa^{(a)}_{Dmm}
\end{pmatrix},\ 
 \bar\kappa^{(b)}_D=
\begin{pmatrix}
  \kappa^{(a)}_{Dii} & {\rho^{(a)}_{Dim}} \cr
  -{\rho^{(a)T}_{Dim}} & \bar\kappa^{(a)}_{Dmm}
\end{pmatrix},
\label{eq:bContfroma}
\end{equation}
where $1_{ii}$ is the $N \times N$ unit matrix.
These basic contractions can be calculated also within the $P$ space.
Thus, the contractions for the original basis are obtained
as in Eq.~(\ref{eq:cbDcontrn}) in the previous subsection,
and so are the overlaps of arbitrary observables;
i.e., most of their calculations can be performed within the $P$ space.

Now we discuss the method to further
reduce the calculation by taking account of the core contributions,
where the core means the subspace composed of the canonical orbits
which have almost full occupation probability, $v^2\approx 1$,
(deep hole states). 
More precisely, setting up a small number $\epsilon$,
we select the core space $O$ composed of
the canonical orbits which satisfy $u^2_k=1-v^2_k < \epsilon$,
$k=1,2,...,L_o(\epsilon)$, for $|\Phi\rangle$
and, $u'^2_k=1-v'^2_k < \epsilon$,
$k=1,2,...,L_{o'}(\epsilon)$, for $|\Phi'\rangle$, respectively,
in a similar manner as selecting the $P$ space.
Namely, the p-h bases satisfy
(omitting $(\epsilon)$ in $L_o(\epsilon)$ and $L_{o'}(\epsilon)$)
\begin{equation}
 \hat a_k|\Phi\rangle=0,\quad k \le L_o \,, \qquad
 \hat a'_{k'}|\Phi'\rangle=0,\quad k' \le L_{o'}.
\label{eq:actrn}
\end{equation}
Note that the core subspace $O$ is contained in the $P$ space,
$P$=$O \oplus \bar P$, and inequalities $0 \le L_o \le N \le L_p \le M$
and $0 \le L_{o'} \le N \le L_{p'} \le M$ hold.
The dimensions of the non-zero Thouless amplitudes for the p-h
bases in Eq.~(\ref{eq:Thph}) are then further reduced,
\begin{equation}
 Z_{pp}= \begin{pmatrix}
  0 & 0 \cr
  0 & Z_{\bar p \bar p}
\end{pmatrix},\qquad
 Z^\prime _{p'p'}= \begin{pmatrix}
  0 & 0 \cr
  0 & Z^\prime _{\bar p' \bar p'}
\end{pmatrix},
\label{eq:Zotran}
\end{equation}
and then
\begin{equation}
 Z'_{Dpp}=(X_D^{-\dagger})_{p \bar p'}Z'_{\bar p' \bar p'}
 \left[1+S_{D \bar p' \bar p'}^\dagger Z'_{\bar p' \bar p'}\right]^{-1}
 (X_D^{-*})_{\bar p'p}+{\tilde S}_{Dpp},
\label{eq:ZDpo}
\end{equation}
where the submatrix $(X_D^{-\dagger})_{p \bar p'}$ is defined by
\begin{equation}
(X_D^{-\dagger})_{p \bar p'}\equiv
((X_D^{-\dagger})_{kk'}; k=1,2,...,L_p,k'=L_{o'}+1,L_{o'}+2,...,L_{p'}),
\end{equation}
and the sizes of square submatrices
$Z_{\bar p \bar p}$ and $Z_{\bar p' \bar p'}$ in the $\bar P$ space
are $L_{\bar p}\equiv L_p-L_o$ and $L_{\bar p'}\equiv L_{p'}-L_{o'}$,
respectively.  Then the calculation of the norm overlap
in Eq.~(\ref{eq:DPhiovlp}) is further reduced in such a way
that the determinants or the pfaffians have smaller sizes
$L_p \rightarrow L_{\bar p}$ and $L_{p'} \rightarrow L_{\bar p'}$.
As for the contractions, although the reductions of the dimensions
of matrix manipulation are restrictive, their effect is still considerable.

From Eq.~(\ref{eq:Zotran}), the reduced contraction $\bar\kappa^{(a)}_{Dpp}$ 
has a subblock form,
\begin{equation}
 \bar\kappa^{(a)}_{Dpp} =
\begin{pmatrix}
  0 & 0 \cr
  0 & [1+Z^\dagger_{\bar p \bar p} Z'_{D\bar p \bar p}]^{-1}
 Z_{\bar p \bar p}^\dagger
\end{pmatrix} \equiv
\begin{pmatrix}
  0 & 0 \cr
  0 & \bar\kappa^{(a)}_{D\bar p \bar p}
\end{pmatrix},
\end{equation}
and then
\begin{equation}
 \rho^{(a)}_{Dp'p} =
\begin{pmatrix}
  0 & Z'_{D o \bar p} \bar\kappa^{(a)}_{D\bar p \bar p} \cr
  0 & Z'_{D \bar p' \bar p} \bar\kappa^{(a)}_{D\bar p \bar p}
\end{pmatrix},\qquad
 \kappa^{(a)}_{Dp'p'} =
Z'_{D p'p'}
-Z'_{D p' \bar p} \bar\kappa^{(a)}_{D\bar p \bar p} Z'_{D \bar p p'},
\end{equation}
where the subblock matrices of $Z'_D$ are defined obviously by
\begin{equation}
\begin{array}{ll}
 Z'_{Do\bar p} &\equiv
 \left((Z'_D)_{kk'};k=1,2,...,L_o,k'=L_o+1,L_o+2,...,L_p\right),\cr
 Z'_{D\bar p'\bar p} &\equiv
 \left((Z'_D)_{kk'};k=L_o+1,L_o+2,...,L_{p'},k'=L_o+1,L_o+2,...,L_p\right),\cr
 Z'_{Dp'\bar p} &\equiv
 \left((Z'_D)_{kk'};k=1,2,...,L_{p'},k'=L_o+1,L_o+2,...,L_p\right),\cr
 Z'_{D\bar p p'} &\equiv
 \left((Z'_D)_{kk'};k=L_o+1,L_o+2,...,L_p,k'=1,2,...,L_{p'}\right).
\end{array}
\end{equation}
Namely, non-zero part of $\bar\kappa^{(a)}_D$ is reduced
from $L_p\times L_p$ to $L_{\bar p}\times L_{\bar p}$,
that of $\rho^{(a)}_{Dp'p}$
from $L_{p'}\times L_p$ to $L_{p'}\times L_{\bar p}$,
while that of $\kappa^{(a)}_D$ is unchanged and $L_{p'}\times L_{p'}$.
Using these contractions and Eq.~(\ref{eq:bContfroma}),
the basic contractions for the $(\hat b^\dagger,\hat b)$ basis
take the following subblock forms,
\begin{equation}
 \rho^{(b)}_{Dp'p}=
\begin{pmatrix}
  1_{oo} & 0 \cr
  \rho^{(b)}_{D\bar p'o} & \rho^{(b)}_{D\bar p' \bar p}
\end{pmatrix},\quad
 \kappa^{(b)}_{Dp'p'}=
\begin{pmatrix}
  0 & 0 \cr
  0 & \kappa^{(b)}_{D\bar p' \bar p'}
\end{pmatrix},\quad
 \bar\kappa^{(b)}_{Dpp}=
\begin{pmatrix}
  \bar\kappa^{(b)}_{Doo} & \bar\kappa^{(b)}_{Do \bar p} \cr
  \bar\kappa^{(b)}_{D\bar p o} & \bar\kappa^{(b)}_{D\bar p \bar p}
\end{pmatrix},
\end{equation}
where $1_{oo}$ is the $L_o\times L_o$ unit matrix.
Thus, the overlap calculations of
one-body and two-body operators in Eqs.~(\ref{eq:FGbtrn}),
(\ref{eq:FFtrn}), and~(\ref{eq:GGtrn}) are considerably reduced,
especially for heavy nuclei with weak pairing correlations.

In this way, we have shown that the truncation scheme
within the $P$ space works
for more general representations based on the p-h vacuums
(Slater determinants),
although the formula are more complicated.
Furthermore, the additional reduction of matrix manipulations is
possible related to the core contributions.
Various subblocks for the matrix representation of the amplitudes
or of observables in the $(\hat b^\dagger,\hat b)$
or $(\hat a^\dagger,\hat a)$ basis are introduced;
the $P$ and $Q$ spaces, the particle and hole spaces,
and the core space $O$ with $P=O\oplus \bar P$.
They are summarized for the Thouless amplitude $Z$
for the $(\hat a^\dagger,\hat a)$ basis as
\begin{equation}
 Z = \bordermatrix{
  & i_o & i_{\bar p} & m_{\bar p} & m_q \cr
  i_o & 0 & 0 & 0 & 0 \cr
  i_{\bar p} & 0 & * & * & 0 \cr
  m_{\bar p} & 0 & * & * & 0 \cr
  m_q & 0 & 0 & 0 & 0 },
\end{equation}
where the subindex $i_o$ denotes the core orbits,
$i_{\bar p}$ the remaining hole orbits,
$m_{\bar p}$ the particle orbits in the $P$ space,
and $m_q$ the $Q$ space orbits.
Their borders are specified by the dimensions,
$L_o$, $N$ (particle number), and $L_p$, respectively,
in the full space dimension $M$.

Finally we mention that
the arbitrary phases of the HFB type states in Eq.~(\ref{eq:Zanor}) can be
conveniently chosen;
\begin{equation}
 n=|\det U_a|^{1/2}
 =\det\left[1+Z^\dagger Z\right]^{-1/4}_{\bar p \bar p}, \quad
 n'=|\det U'_{a'}|^{1/2}
 =\det\left[1+Z'^\dagger Z'\right]^{-1/4}_{\bar p' \bar p'},
\end{equation}
which naturally guarantee the condition,
$|\Phi\rangle \rightarrow |\phi\rangle$ and
$|\Phi'\rangle \rightarrow |\phi'\rangle$
in the limit of vanishing pairing correlations.
In this limit, the basic contractions take the forms
\begin{equation}
 \rho^{(b)}_{Dp'p}\rightarrow 1_{ii},\qquad
 \kappa^{(b)}_{Dp'p'}\rightarrow 0, \qquad 
 \bar\kappa^{(b)}_{Dpp}\rightarrow 0,
\label{eq:bConlim}
\end{equation}
with which the formula for the quantum number projection
(and/or the configuration mixing) for the Slater determinantal states
are recovered.

\section{Example calculations}
\label{sec:example}

\subsection{Choice of Hamiltonian}
\label{sec:hamil}

In this section, we show some examples of the result of calculations,
which are obtained by applying the formulation developed
in the previous section.
It is required to start with the spherically invariant two-body Hamiltonian.
Although it is desirable to use realistic interactions like
the Gogny or Skyrme forces, it has been recognized that
the density-dependent part of interaction causes some problems
for the quantum number projection and/or the configuration mixing
calculations; see e.g. Refs.~\cite{Rob07,DSN07,BDL09}.
In this paper, we restrict ourselves
to the schematic multipole-multipole two-body interactions for simplicity.
However, in order to make the result as realistic as possible,
we employ the Woods-Saxon potential as a mean-field, and
construct the residual multipole interactions consistent with it
according to Ref.~\cite{BM75}.
Needless to say, the Hamiltonian is spherical invariant;
therefore, we start from a hypothetical spherical ground state
for the construction.

Thus, our Hamiltonian is written as
\begin{equation}
 \hat H=\hat h + \hat H_F + \hat H_G,\qquad
 \hat h=\hat h_0+\hat h_1,\qquad
 \hat h_0=\sum_{\tau={\rm n,p}}\left(\hat t_\tau+\hat V^\tau_{\rm WS}\right),
\label{eq:hamT}
\end{equation}
where $\hat h_0$ is a spherical mean-field Hamiltonian composed
of the kinetic energy and the Woods-Saxon potential (with the Coulomb
interaction for proton), and $\tau={\rm n,p}$ distinguishes neutron or proton.
The part $\hat h_1$ is included to cancel out the exchange contributions
coming from the residual interactions $\hat H_F$ and $\hat H_G$,
and is discussed later.
Assuming the same spatial deformation for neutron and proton,
the particle-hole type ($F$-type) isoscalar interaction $\hat H_F$
is given by
\begin{equation}
 \hat H_F=-\frac{1}{2}\,\chi\sum_{\lambda\ge 2}
 \sum_\mu (-1)^\mu :\hat F_{\lambda-\mu} \hat F_{\lambda\mu}:,
 \qquad \hat F_{\lambda\mu}=\sum_{\tau={\rm n,p}} \hat F_{\lambda\mu}^\tau,
\label{eq:hamF}
\end{equation}
where the operator $\hat F^\tau_{\lambda\mu}$,
\begin{equation}
 \hat F^\tau_{\lambda\mu}\equiv
 \sum_{ij} \langle i| \hat F^\tau_{\lambda\mu}|j \rangle
 \hat c^\dagger_i \hat c_j,
\label{eq:opF}
\end{equation}
is defined by the one-body field,
\begin{equation}
 F^\tau_{\lambda\mu}(\bm{r})=R^\tau_0\, \frac{dV_c^\tau}{dr}\,
 Y_{\lambda\mu}(\theta,\phi),
\label{eq:hamFop}
\end{equation}
with $V_c^\tau(r)$ and $R^\tau_0$ being the central part of
the Woods-Saxon potential and its radius, respectively.
As is already mentioned,
we employ the spherical harmonic oscillator basis
as the original basis states \{$|i \rangle$\}.
The selfconsistent force parameter $\chi$ is independent of
the multipolarity $\lambda$ and is given by
\begin{equation}
 \chi=(\kappa_n+\kappa_p)^{-1},\qquad
 \kappa_\tau\equiv \left(R_0^\tau\right)^2
 \int \rho_0^\tau(r)\frac{d}{dr}\left(r^2\frac{dV^\tau_c(r)}{dr}\right) dr,
\label{eq:strF}
\end{equation}
where
$\rho_0^\tau(r)$ is the spherical ground state density,
which is calculated with the filling approximation for each nucleus
based on the spherical Woods-Saxon single-particle state of $\hat h_0$.

As for the pairing ($G$-type) interaction $\hat H_G$,
\begin{equation}
 \hat H_G=-\sum_{\tau,\lambda\ge 0} g^\tau_\lambda\,
 \sum_\mu\hat G^{\tau\dagger}_{\lambda\mu} \hat G^\tau_{\lambda\mu},\qquad
 \hat G^{\tau\dagger}_{\lambda\mu}\equiv\frac{1}{2}\sum_{ij}
 \langle i| \tilde G^{\tau}_{\lambda\mu} |j \rangle
 \hat c^\dagger_i \hat c^\dagger_{\tilde j},
\label{eq:hamG}
\end{equation}
where $\tilde j$ denotes the time reversal conjugate state of $j$,
we employ the standard multipole form defined by the operator,
\begin{equation}
 \tilde G_{\lambda\mu}(\bm{r})=\left(\frac{r}{\bar R_0}\right)^\lambda
 \sqrt{\frac{4\pi}{2\lambda+1}}\, Y_{\lambda\mu}(\theta,\phi),
\label{eq:opG}
\end{equation}
with $\bar R_0=1.2 A^{1/3}$ fm.
Just like the zero-range interactions, this type of
simplified pairing interactions cannot be used with the full model space.
We employ cut-off of the matrix elements
for the operator $\tilde G^\tau_{\lambda\mu}$; namely the following
replacement is done:
\begin{equation}
 \langle i | \tilde G_{\lambda\mu}^\tau |j \rangle
 \rightarrow
 \sum_{kl} w^{0\tau *}_{ik} w^{0\tau}_{jl}
 \left[f_{\rm c}(\epsilon_k^{0\tau}) f_{\rm c}(\epsilon_l^{0\tau})\right]^{1/2}
 \times \langle k | \tilde G_{\lambda\mu}^\tau |l \rangle_{\rm WS}^0,
\label{eq:Gcutoff}
\end{equation}
where $\epsilon_l^{0\tau}$ and
$\langle k | \tilde G_{\lambda\mu}^\tau |l \rangle_{\rm WS}^0$ are
the eigenenergies of the spherical Woods-Saxon states
and the matrix elements with respect to them, respectively,
and $w^{0\tau}_{ik}$ is their transformation matrix
from the original harmonic oscillator basis states.
We use the following form of the cut-off factor~\cite{TST10},
\begin{equation}
f_{\rm c}(\epsilon) =
\frac{1}{2} \left[1+
{\rm erf}\left(\frac{ \epsilon-\lambda+\Lambda\rs{l}}{d\rs{cut}}\right)
\right]^{1/2}
\left[1+
{\rm erf}\left(\frac{-\epsilon+\lambda+\Lambda\rs{u}}{d\rs{cut}}\right)
\right]^{1/2},
\label{eq:fcutoff}
\end{equation}
where the error function is defined by
${\displaystyle
{\rm erf}(x)=\frac{2}{\sqrt{\pi}} \int_{0}^{x}e^{-t^2}dt }$,
and the parameters are chosen to be
$\Lambda\rs{u}=\Lambda\rs{l}=1.2\,\hbar \omega$ and
$d\rs{cut}=0.2\,\hbar \omega$ with $\hbar\omega=41/A^{1/3}$ MeV.
The quantity $\lambda$ in the cut-off factor in Eq.(\ref{eq:fcutoff}) is
the chemical potential determined to guarantee the correct average number
in the treatment of pairing correlation (see the next subsection).

It should be noted that all the two-body terms, including the exchange
contributions, are evaluated in the calculation of
the quantum number projection.
Even for the hypothetical spherical ground state, the exchange term
of the interaction $\hat H_F$ and $\hat H_G$ induces
extra spherical one-body fields,
which are written explicitly as,
\begin{equation}
 \hat h_F\equiv\chi \sum_{\tau,\lambda\ge 2} \sum_{ij}\biggl( \sum_\mu\sum_{kl}
 (-1)^\mu \langle i |\hat F^\tau_{\lambda-\mu}|k\rangle
 (\rho^\tau_0)_{kl}
 \langle l |\hat F^\tau_{\lambda\mu}|j\rangle\biggr)\hat c_i^\dagger c_j,
\label{eq:exchgF}
\end{equation}
and
\begin{equation}
 \hat h_G\equiv-\sum_{\tau,\lambda\ge 0} g^\tau_\lambda \sum_{ij}
 \biggl(\sum_\mu \sum_{kl}
 \langle il|\hat G^{\tau\dagger}_{\lambda\mu}|\rangle
 (\rho^\tau_0)_{kl}
 \langle|\hat G^\tau_{\lambda\mu}|jk\rangle\biggr)\hat c_i^\dagger c_j,
\label{eq:exchgG}
\end{equation}
where $(\rho^\tau_0)_{kl}$ is the density matrix
for the spherical ground state.
Since the Hamiltonian consists of the one-body part and
its residual interaction for the spherical ground state,
we subtract these terms from the one-body Hamiltonian $\hat h_0$ and
the one-body field $\hat h_1$ in Eq.~(\ref{eq:hamT}) is given by
\begin{equation}
 \hat h_1=-\hat h_F-\hat h_G.
\label{eq:h1}
\end{equation}
Note that this term $\hat h_1$ is not used to generate the mean-field
states from which the projection calculations are performed.

As for the parameter set for the Woods-Saxon potential,
we use the one recently proposed by Ramon Wyss~\cite{WyssPrivate05}
and employed in Refs.~\cite{SS09,TST10,MSK11},
which very nicely reproduces the geometrical property like the nuclear radius.

\subsection{Details of calculation}
\label{sec:calc}

We have developed a program to perform the general quantum number projection
and the configuration mixing for the Hamiltonian in Eq.~(\ref{eq:hamT})
according to the formulation in Sec.~\ref{sec:formulation}.
We have made the program in such a way that most general
symmetry-breaking mean-field states (HFB type states)
$|\Phi\rangle$ can be accepted
as long as they are expanded in the spherical harmonic oscillator basis.
More precisely, the angular momentum projection, neutron and proton number
projections, and the parity projection are performed simultaneously;
and, optionally, the configuration
mixing in the sense of the GCM can be done.
Namely, the final nuclear wave function is expressed as,
\begin{equation}
 |\Psi_{M;\alpha}^{INZ(\pm)}\rangle = \sum_{K,n} g_{Kn,\alpha}^{INZ(\pm)}\,
  \hat P_{MK}^I \hat P^N \hat P^Z \hat P_{\pm}|\Phi_n\rangle.
\label{eq:INZpiProj}
\end{equation}
The projectors are given, as usual, by
\begin{equation}
 \hat P_{MK}^I=\frac{2I+1}{8\pi^2}\int d^3\omega
 D_{MK}^{I\,*}(\omega)\hat R(\omega),
 \qquad \hat P^N = \frac{1}{2\pi}\int d\varphi \,e^{i\varphi(\hat N -N)},
\label{eq:ProjOps}
\end{equation}
the similar one for the proton number projector $\hat P^Z$,
and the parity projector $\hat P_{\pm}$ in Eq.~(\ref{eq:ProjP}).
The mixing amplitude $g_{Kn,\alpha}^{INZ(\pm)}$ is obtained by solving
the generalized eigenvalue problem of the Hill-Wheeler equation,
\begin{equation}
 \sum_{K',n'}{\cal H}_{Kn;K'n'}^{INZ(\pm)}\,g_{K'n',\alpha}^{INZ(\pm)}
 =E_\alpha^{INZ(\pm)}
 \sum_{K',n'}{\cal N}_{Kn;K'n'}^{INZ(\pm)}\,g_{K'n',\alpha}^{INZ(\pm)},
\label{eq:HWeq}
\end{equation}
where the Hamiltonian and norm kernels are defined as
\begin{equation}
 \begin{pmatrix}
 {\cal H}_{Kn;K'n'}^{INZ(\pm)} \cr
 {\cal N}_{Kn;K'n'}^{INZ(\pm)} \end{pmatrix}
 =\langle \Phi_n|\begin{pmatrix} \hat H \cr 1 \end{pmatrix}
  \hat P_{KK'}^I \hat P^N \hat P^Z \hat P_{\pm}|\Phi_{n'}\rangle.
\label{eq:HNkernels}
\end{equation}
In the present paper, however,
we only show the results of the quantum number projection;
namely, no configuration mixing is performed.

The generalized eigenvalue problem in Eq.~(\ref{eq:HWeq}) is
solved in a standard way, i.e., the so-called two step method.
Namely, first the norm kernel is diagonalized and the states
with small norm eigenvalue are discarded, and then the remaining
energy eigenvalue problem is solved in the restricted space.
In the present work for the general quantum number projections,
we have excluded the state whose norm eigenvalue is smaller than $10^{-13}$.
The numerical integrations for the projectors in Eq.~(\ref{eq:ProjOps})
is treated by the standard Gaussian quadratures.
It should be noted that since we do not impose any symmetry like $D_2$
the number of points required for the Gaussian quadratures
are considerably large.

As for the mean-field state $|\Phi\rangle$,
it may be desirable to apply the HFB procedure.
However, we found that the schematic separable type interaction
in Eq.~(\ref{eq:hamT}) with large model space does not always
gives a reasonable result, e.g., the appropriate ground state deformation.
Therefore, in the present work,
we utilize the following deformed mean-field Hamiltonian,
\begin{equation}
 \hat h_{\rm def} = \hat h_0
 -\sum_{\lambda\mu}\alpha_{\lambda\mu}^* \hat F_{\lambda\mu},
\label{eq:defh}
\end{equation}
where the deformation parameters $\{\alpha_{\lambda\mu}\}$ are basically
determined by the Woods-Saxon Strutinsky calculation of Ref.~\cite{TST10}.
The deformed mean-field in Eq.~(\ref{eq:defh}) is obtained
from the schematic interaction~(\ref{eq:hamF})
in the Hartree-Bogoliubov (HB) approximation
if the selfconsistent condition,
$\alpha_{\lambda\mu}=\chi\langle \Phi|\hat F_{\lambda\mu}|\Phi\rangle$,
is satisfied.  At the same time, it coincides, within the first order
in the deformation parameters, with the central part of
the standard deformed Woods-Saxon potential~\cite{CDN87},
which is used in Ref.~\cite{TST10}.
The potential is defined with respect to
the deformed nuclear surface specified by the radius,
\begin{equation}
 R(\theta,\phi)=R_0\, c_v(\{\alpha_{\lambda\mu}\})
 \left[1+\sum_{\lambda\mu}\alpha^*_{\lambda\mu}
Y_{\lambda\mu}(\theta,\phi)\right],
\label{eq:Rdef}
\end{equation}
where the constant $c_v(\{\alpha_{\lambda\mu}\})$ takes care of
the volume conservation.

With the deformed Hamiltonian in Eq.~(\ref{eq:defh}),
the mean-field state $|\Phi\rangle$ is generated by the paired and
cranked mean-field,
\begin{equation}
 \hat h'_{\rm mf}= \hat h_{\rm def}
 -\sum_{\tau={\rm n,p}}{\mit\Delta}_\tau
  \left(\hat P_\tau^\dagger+\hat P_\tau \right)
 -\sum_{\tau={\rm n,p}}\lambda_\tau \hat N_\tau
 -\omega_{\rm rot} \hat J_x,
\label{eq:cpdefh}
\end{equation}
where $\hat N_\tau$ and $\lambda_\tau$ are the particle number operator
and the chemical potential, respectively, while
$\hat P^\dagger_\tau=\hat G_{00}^{\tau\dagger}$ and
${\mit\Delta}_\tau = g_0^\tau \langle \Phi | \hat G^\tau_{00} |\Phi \rangle$,
namely the static monopole pairing part in the Hamiltonian
in Eq.~(\ref{eq:hamG}) is included selfconsistently
within the HB procedure to generate the mean-field state $|\Phi \rangle$.
The ground states of nuclei considered in the present example calculations
are axially symmetric, $\alpha_{\lambda\mu}=0$ for $\mu \ne 0$,
and the effect of the rotation about the $x$-axis perpendicular to
the symmetry axis is taken into account
with the rotational frequency $\omega_{\rm rot}$.
We do not intend to study high spin states in the present work,
and are mainly concerned about the ground state band.
However, we found that the $K$ mixing caused by the cranking procedure
is essential to reproduce the moment of inertia; as will be discussed
in the followings, a small cranking frequency is enough for such a purpose.

By using the Woods-Saxon Strutinsky calculation of Ref.~\cite{TST10}
the axially symmetric quadrupole and hexadecapole deformations,
$\alpha_{20}$ and $\alpha_{40}$, are determined.
For parity breaking case, we additionally include $\alpha_{30}$
deformation in such a way to roughly reproduce the energy splitting
of the ground state parity doublet bands.  Correspondingly,
we include $\lambda=2,3,4$ components in the isoscalar $F$-type
interactions in Eq.~(\ref{eq:hamF}) with the common selfconsistent 
strength $\chi$ given in Eq.~(\ref{eq:strF}).  As for the $G$-type interaction,
we include $\lambda=0,2$ components.  The monopole pairing ($\lambda=0$)
strength $g_0^\tau$ is determined so that the pairing gap ${\mit\Delta}_\tau$
at zero frequency $\omega_{\rm rot}=0$
in Eq.~(\ref{eq:cpdefh}) reproduces the even-odd mass differences.
The quadrupole pairing strength is assumed to be proportional
to the monopole pairing strength and the proportionality constant,
which is assumed to be common to neutron and proton,
is chosen to reproduce the final rotational spectra.
We assume the constant deformations for the cranking calculation
for simplicity.  The effects of cranking for the results of
angular-momentum-projection calculation
are discussed in the following two examples.

\subsection{Rotational spectrum in $^{\bf 164}$Er}
\label{sec:Er}

As a first example, we consider a typical rotational spectrum
of the ground state band in the rare earth region,
taking a nucleus $^{164}$Er.
The parameters determined according to the procedure explained
in the previous subsection and used in the calculation are
summarized in Table~\ref{tab:Er}.
Strictly speaking, the values of the $F$-type interaction strength $\chi$
and the monopole pairing interaction strength $g_0^\tau$ depend on
the size of the spherical oscillator basis.
However, their dependences are very weak if the size is large enough.
We present the values for $\noscmax=18$.
In this case the parity of the mean-field is conserved and
the parity projection is unnecessary;
all the states belong to the positive parity.
To perform the angular momentum projection,
the numbers of points for the Gaussian quadratures
with respect to the Euler angles, $\omega=(\alpha,\beta,\gamma)$,
are $N_\alpha=N_\gamma=16$ and $N_\beta=50$ for the non-cranked case
and for the case with the small cranking frequency
$\hbar\omega_{\rm rot}=0.01$ MeV.
For the cases with larger cranking frequencies, they are increased to
$N_\alpha=N_\gamma=22$ and $N_\beta=70$.
As for the number projection, the number of mesh points with respect to
the gauge angle is $N_\varphi=17$ for both neutron and proton.
These values are chosen to guarantee the convergence of the results.

\begin{table}[ht]
\begin{center}
\begin{tabular}{ccccccccc}
\hline
$\alpha_{20}$ & $\alpha_{30}$ & $\alpha_{40}$ & $\chi$ [MeV$^{-1}$] &
${\mit\Delta}_{\rm n}$ [MeV] & ${\mit\Delta}_{\rm p}$ [MeV] & 
$g_0^{\rm n}$ [MeV] & $g_0^{\rm p}$ [MeV] & $g_2^\tau/g_0^\tau$ \cr
\hline
0.276\  & 0\  & 0.012\  & $2.566\times10^{-4}$ & 1.020 & 1.025 &
0.1606 & 0.2096 & 13.60 \cr
\hline
\end{tabular}
\end{center}
\caption{
The parameters used in the calculation for $^{164}_{\ 68}$Er$_{96}$.
The values of $\chi$ and $g_0^\tau$ are those with the size of
basis $\noscmax=18$.
}
\label{tab:Er}
\end{table}

\begin{figure}[!ht]
\begin{center}
\includegraphics[width=150mm]{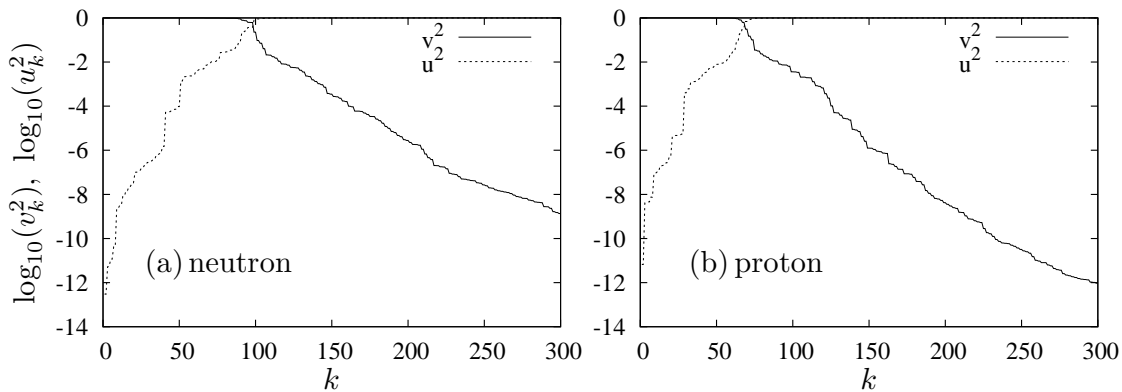}
\vspace*{-5mm}
\caption{
Occupation and empty probabilities $v_k^2$ and $u_k^2=1-v_k^2$
as functions of the number $k$ of the canonical basis for $^{164}$Er;
the log scale is used for the abscissa.
The panel (a) is for neutron and (b) for proton.
The spherical oscillator shells $\noscmax=18$ is used.
}
\label{fig:Ervvuu}
\end{center}
\end{figure}

\begin{figure}[!ht]
\begin{center}
\includegraphics[width=130mm]{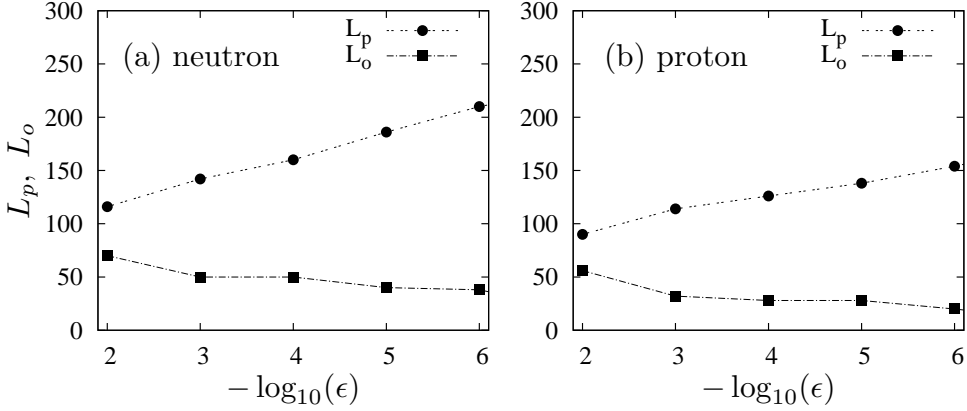}
\vspace*{-5mm}
\caption{
The number of levels in the model space ($P$-space) $L_p$
and the the number of core levels $L_o$ as functions of
the small number $\epsilon$ for $^{164}$Er;
the log scale is used for the ordinate.
The panel (a) is for neutron and (b) for proton.
The spherical oscillator shells $\noscmax=18$ is used.
}
\label{fig:ErepsLp}
\end{center}
\end{figure}

First of all, we show the occupation probability of the canonical basis
in Fig.~\ref{fig:Ervvuu} in the logarithmic scale, which is a measure
how important each canonical orbit is.
Not only the occupation probability $v^2_k$ but the empty probability
$u^2_k=1-v^2_k$ are shown.
The quantity $v^2_k$ quickly decreases after $k>N=96$ for neutron
and $k>Z=68$ for proton.  The truncation of the model space is based
on the smallness of the occupation probability as is explained
in \S~\ref{sec:trunc}.
On the other hand, the quantity $u^2_k$ tells how important
the pairing correlation is for deep hole states.
As explained in \S~\ref{sec:truncphv} a part of calculations
can be simplified for the orbits with small $u^2_k$.
As explained in details in \S~\ref{sec:formulation},
the projection calculation is composed of many matrix manipulations.
The dimension of the matrices are determined by the model space
truncation and, partly, by excluding the core contributions.
The sizes of the model space $L_p(\epsilon)$, the number of orbits $k$ which
satisfies $v^2_k < \epsilon$ defined in \S~\ref{sec:trunc},
and the size of the core space $L_o(\epsilon)$, the number of orbits $k$ which
satisfies $u^2_k < \epsilon$ defined in \S~\ref{sec:truncphv}
are presented in Fig.~\ref{fig:ErepsLp}.
As it is clear from the figure, the number $L_p(\epsilon)-L_o(\epsilon)$
is very small compared to, e.g., the number $M=2660$
corresponding to the full size of the oscillator space with $\noscmax=18$.

\begin{figure}[!ht]
\begin{center}
\includegraphics[width=150mm]{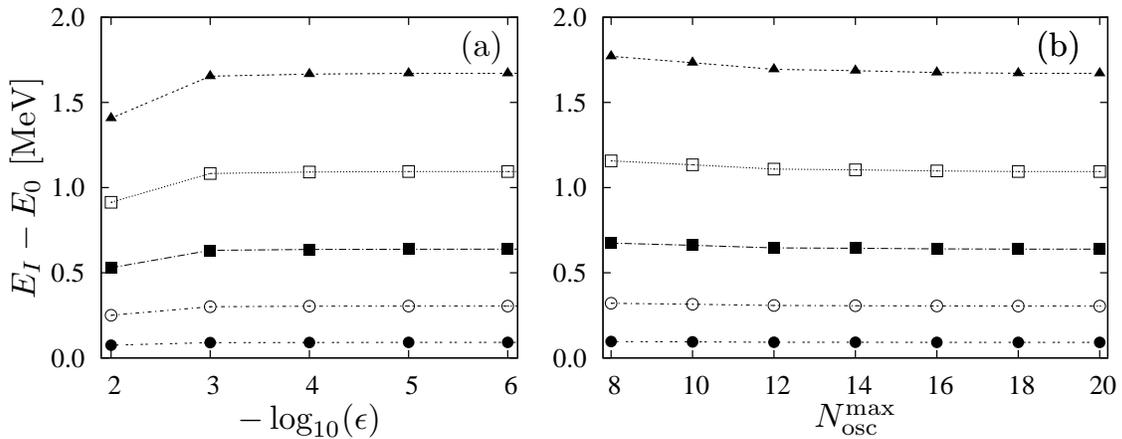}
\vspace*{-5mm}
\caption{
The rotational excitation spectra from $I=2$ to~8
calculated by the angular momentum projection for $^{164}$Er
as functions of the cut-off parameter $\epsilon$ (panel (a))
with $\noscmax=18$,
and of the size of the spherical harmonic oscillator basis $\noscmax$
(panel (b)) with $\epsilon=10^{-6}$.
The cranking frequency $\hbar\omega_{\rm rot}=0.01$ MeV is used.
}
\label{fig:Erconv}
\end{center}
\end{figure}

In order to see how the truncated model space can be chosen,
we show in the left panel of Fig.~\ref{fig:Erconv} the final rotational
spectra as functions of the cut-off parameter $\epsilon$.
It is clear that $\epsilon \approx 10^{-4}-10^{-5}$ is enough
to obtain the convergent results.
If we take $\epsilon \approx 10^{-4}$, $L_p \approx 160$ (130)
and $L_o \approx 50$ (30) for neutron (proton).
Therefore the size of reduction of model space from $M=2660$
($\noscmax=18$) to $L_p-L_o$ is about factor 25 in this case.
In the right panel of Fig.~\ref{fig:Erconv}
the convergence of the same rotational spectra
with respect to the size of the spherical harmonic oscillator space,
$\noscmax$, is shown.
The basis truncation of $\noscmax=10-12$ has been done sometimes
for the mean-field calculations.
However, the convergence is not enough for higher spin states,
and the larger size $\noscmax\approx 18$ is necessary for obtaining
the stable excitation energy for the $I=8$ member.
In the following the results with $\epsilon=10^{-6}$, $\noscmax=18$
and $\hbar\omega_{\rm rot}=0.01$ MeV are shown if the values of them
are not explicitly mentioned.

\begin{figure}[!ht]
\begin{center}
\includegraphics[width=150mm]{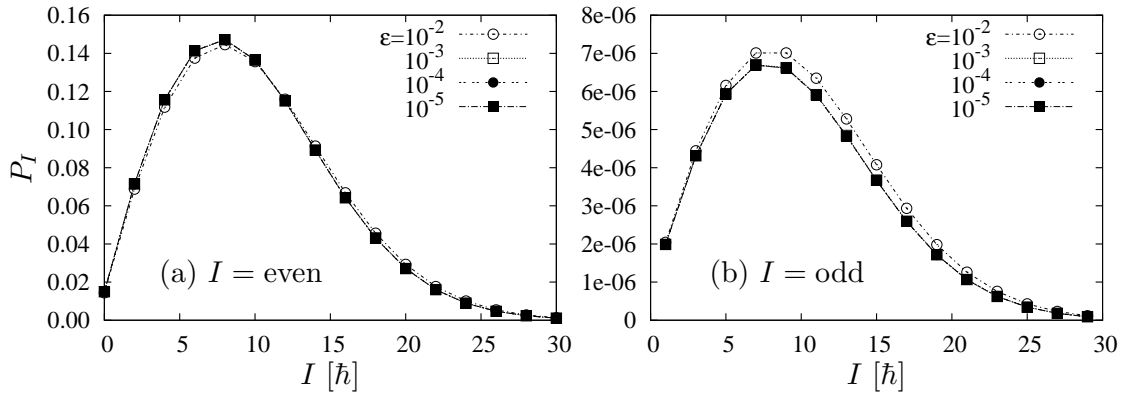}
\vspace*{-5mm}
\caption{
The $I$ distribution of the mean-field state for $^{164}$Er.
The cranking frequency is $\hbar\omega_{\rm rot}=0.01$ MeV.
Even and odd $I$ distributions are plotted separately,
because the absolute values are very different.
Four cases with different values of the cut-off parameter $\epsilon$
are included.
}
\label{fig:ErIdst}
\end{center}
\end{figure}

\begin{figure}[!ht]
\begin{center}
\includegraphics[width=80mm]{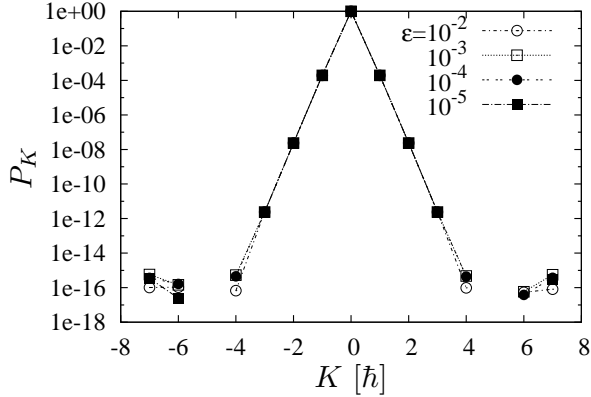}
\vspace*{-5mm}
\caption{
The $K$ distribution of the mean-field state for $^{164}$Er;
the log scale is used for the abscissa.
The cranking frequency is $\hbar\omega_{\rm rot}=0.01$ MeV.
Four cases with different values of the cut-off parameter $\epsilon$
are included.
}
\label{fig:ErKdst}
\end{center}
\end{figure}

\begin{figure}[!ht]
\begin{center}
\includegraphics[width=150mm]{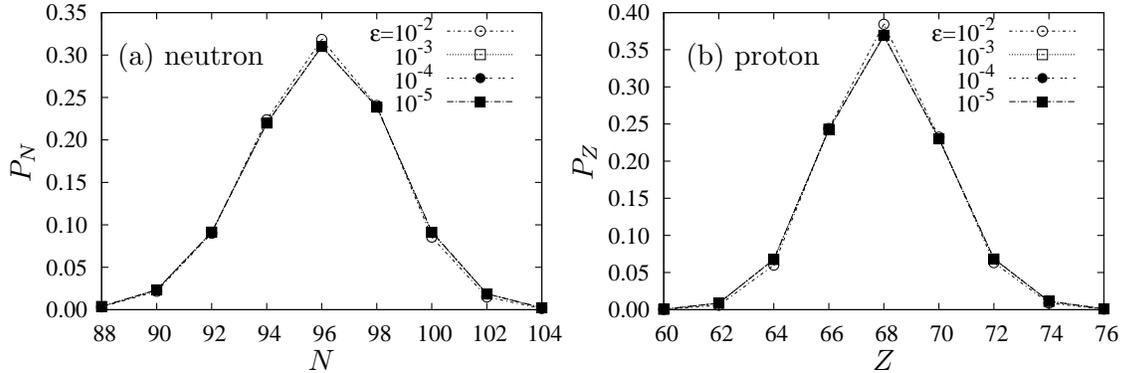}
\vspace*{-5mm}
\caption{
The number distributions of the mean-field state for $^{164}$Er,
the panel (a) is for neutron and (b) for proton.
Four cases with different values of the cut-off parameter $\epsilon$
are included.
}
\label{fig:ErNdst}
\end{center}
\end{figure}

Although it is already rather well-known, we show in Fig.~\ref{fig:ErIdst}
the distribution of the angular momentum $I$ in the mean-field state
$|\Phi\rangle$, namely,
\begin{equation}
 P_I\equiv \sum_{K} \langle\Phi|\hat P^{I}_{KK} |\Phi\rangle
 /\langle\Phi|\Phi\rangle.
\end{equation}
The results with several $\epsilon$ values are also included:
Again $\epsilon=10^{-4}$ is enough for converged results.
It is noticed that the probability of having the odd $I$ components 
is non-zero because the mean-field state is cranked
with small frequency $\omega_{\rm rot}=0.01$ MeV.
If is used the non-cranked state, the odd $I$ components
are strictly zero because of the signature symmetry
(invariance of the $\pi$ rotation about the cranking axis)
and time reversal symmetry
present in the axially symmetric mean-field state.
Next, the distribution of the $K$ quantum number
is shown in Fig.~\ref{fig:ErKdst}:
\begin{equation}
 P_K\equiv \sum_{I} \langle\Phi|\hat P^{I}_{KK} |\Phi\rangle
 /\langle\Phi|\Phi\rangle.
\end{equation}
The $K$ mixing in the wave function
is also due to the Coriolis coupling caused by the cranking procedure;
namely the distribution has only $K=0$ component if the non-cranked
mean-field state is used.
Since the cranking frequency is small $\hbar\omega_{\rm rot}=0.01$ MeV,
the distribution of $K$ is almost linear in $|K|$ in the logarithmic scale.
Although the mixing of $K$ is very small, it is shown that this
${\mit\Delta}K=\pm 1$ mixing is very important
to obtain the proper value of the moment of inertia.
For completeness, we also show the particle number distributions related to
the number projection in Fig.~\ref{fig:ErNdst};
\begin{equation}
 P_N\equiv \langle\Phi|\hat P^N |\Phi\rangle
 /\langle\Phi|\Phi\rangle,\qquad
 P_Z\equiv \langle\Phi|\hat P^Z |\Phi\rangle
 /\langle\Phi|\Phi\rangle.
\end{equation}
The main component corresponding to the correct neutron or proton number
has about 30$-$40\% probability, which is known to be rather standard
for the pairing model space employed presently
and for the typical pairing gap $\Delta \approx 1$ MeV.

\begin{figure}[!ht]
\begin{center}
\includegraphics[width=80mm]{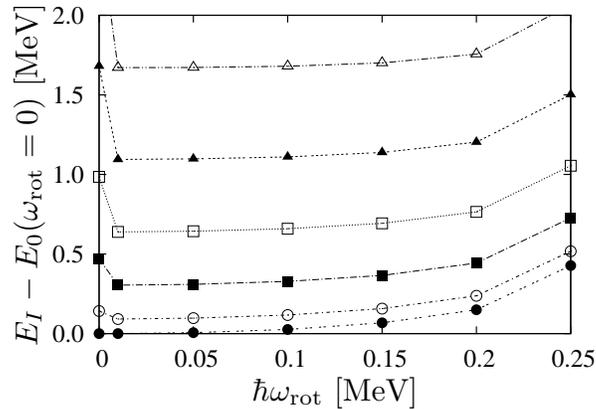}
\vspace*{-5mm}
\caption{
The rotational excitation spectra obtained by the angular momentum
projection from the cranked mean-field state $|\Phi(\omega_{\rm rot})\rangle$
in $^{164}$Er.
}
\label{fig:Errotsp}
\end{center}
\end{figure}

Now we discuss the effect of the cranking on the spectra obtained
by the angular momentum projection.
The cranking procedure is an efficient method to study the high spin
properties of atomic nuclei.
However, we concentrate in this paper on the most fundamental
rotational spectra, i.e., those of the ground state band.
Therefore we only consider the small cranking frequency so that
the two quasiparticle alignment does not occur.
We present the resultant spectra obtained by the angular momentum
projection from the cranked mean-field state
with the frequency $\omega_{\rm rot}$ in Fig.~\ref{fig:Errotsp}.
It is clear that the effect of cranking is very regular and
all the energy $E_I(\omega_{\rm rot})$ with $I=0,2,...,10$ increases gradually.
However, the excitation spectra $E_I(\omega_{\rm rot})-E_0(\omega_{\rm rot})$
is essentially identical at least
in the range $0 < \hbar\omega_{\rm rot} < 0.2$ MeV.
This indicates that all the cranked mean-field states
with $0 < \hbar\omega_{\rm rot} < 0.2$ MeV are roughly equivalent
to generate the ground state rotational band.
We would like to stress that the state with $\omega_{\rm rot}=0$ does not share
this feature; apparently there are discontinuities in the spectra
in Fig.~\ref{fig:Errotsp}, namely
${\rm lim}\,E_I(\omega_{\rm rot}\rightarrow 0) \ne E_I(\omega_{\rm rot}= 0)$.
Note that the moment of inertia of the first $2^+$ state,
$3/(E_2(\omega_{\rm rot})-E_0(\omega_{\rm rot}))$,
takes a value $32.9$ $\hbar^2/$MeV, while the corresponding value
for the non-cranked axially symmetric (only $K=0$) mean-field
is $21.3$ $\hbar^2/$MeV, which is much smaller.
Therefore the ${\mit\Delta}K=\pm 1$ Coriolis coupling effect
in the wave function is very important to increase the moment of inertia.
It has been known that the cranked mean-field is obtained approximately
by the variation after angular momentum projection~\cite{RS80}.
Therefore, the cranking procedure is a simple and efficient way
to recover the correct moment of inertia even with the angular momentum
projection.

The discontinuity of the spectra obtained by projection from
the cranked and non-cranked spectra can be traced back to
the general eigenvalue problem in Eq.~(\ref{eq:HWeq});
discarding the other projectors and the configuration mixing,
it reads, for eigenvalue $E_I$,
\begin{equation}
 \det{\left({\cal H}_{KK'}^I -E_I {\cal N}_{KK'}^I\right)}=0,
\label{eq:HWeq1}
\end{equation}
with
\begin{equation}
 \begin{pmatrix}
 {\cal H}_{KK'}^I \cr
 {\cal N}_{KK'}^I \end{pmatrix}
 =\langle \Phi|\begin{pmatrix} \hat H \cr 1 \end{pmatrix}
  \hat P_{KK'}^I |\Phi\rangle.
\label{eq:HNkernels1}
\end{equation}
In the case of the axially symmetric even-even nuclei,
the signature is a good quantum number, and the following reduced kernels
can be used with restriction $K,K'\ge 0$,
\begin{eqnarray}
 \begin{pmatrix}
 \widetilde {\cal H}_{KK'}^I \cr
 \widetilde {\cal N}_{KK'}^I
 \end{pmatrix}
 &\equiv& \frac{1}{2\sqrt{(1+\delta_{K0})(1+\delta_{K'0})}} \cr
 &&\times
 \begin{pmatrix}
 {\cal H}_{KK'}^I+(-1)^I {\cal H}_{K,-K'}^I
 +(-1)^I {\cal H}_{-K,K'}^I+ {\cal H}_{-K,-K'}^I
 \cr
 {\cal N}_{KK'}^I+(-1)^I {\cal N}_{K,-K'}^I
 +(-1)^I {\cal N}_{-K,K'}^I+ {\cal N}_{-K,-K'}^I
 \end{pmatrix},
\label{eq:HNkernels1r}
\end{eqnarray}
namely, the dimension of the generalized eigenvalue problem is then $(I+1)$
in place of $(2I+1)$.  Now let us consider the problem in the perturbation
theory with respect to the rotational frequency $\omega_{\rm rot}$.
Taking into account the fact that the mean-field state at $\omega_{\rm rot}=0$
has only $K=0$ component, the cranked state is expanded as
in the following,
\begin{eqnarray}
 |\Phi(\omega_{\rm rot})\rangle &=& |\Phi_0(K=0)\rangle
 +\omega_{\rm rot} \bigl( |\Phi_1(K=+1)\rangle+ |\Phi_1(K=-1)\rangle \bigr) \cr
 &+&\omega_{\rm rot}^2 \bigl( |\Phi_2(K=+2)\rangle
      + |\Phi_2(K=0)\rangle + |\Phi_2(K=-2)\rangle \bigr)
 + ... ,
\label{eq:crpert}
\end{eqnarray}
where $|\Phi_0(K=0)\rangle\equiv|\Phi(\omega_{\rm rot}=0)\rangle$,
and so are the reduced kernels,
\begin{equation}
 \widetilde{\cal H}_{KK'}^I -E_I \widetilde {\cal N}_{KK'}^I
 =\sum_{n=0,2,4,...}\omega_{\rm rot}^{K+K'+n}\Bigl(
  h_{KK'}^{I(n)}-E_I n_{KK'}^{I(n)}\Bigr),\quad (K,K'\ge 0),
\label{eq:kerpert}
\end{equation}
with $O(1)$ quantities $h_{KK'}^{I(n)}$ and $n_{KK'}^{I(n)}$.
Because of this $\omega_{\rm rot}$ dependence, it can be easily
confirmed that the determinant can be written as
\begin{equation}
 \det{\left(\widetilde{\cal H}_{KK'}^I -E_I \widetilde {\cal N}_{KK'}^I
 \right)}=
 \omega_{\rm rot}^{2(I+1)}
 \det{\left(h_{KK'}^{I(0)} -E_I n_{KK'}^{I(0)}\right)}
 +O(\omega_{\rm rot}^{2(I+2)}),
\label{eq:genpart}
\end{equation}
and the eigenvalue equation in Eq.~(\ref{eq:HWeq1}) reduces,
in the limit $\omega_{\rm rot}\rightarrow 0$, to
\begin{equation}
 \det{\left(h_{KK'}^{I(0)} -E_I n_{KK'}^{I(0)}\right)}=0.
\label{eq:HWeq2}
\end{equation}
In contrast, if we put $\omega_{\rm rot}=0$
beforehand in Eq.~(\ref{eq:kerpert}),
only $K=K'=0$ kernels survives, and we obtain simply the equation,
\begin{equation}
 h_{00}^{I(0)} -E_I n_{00}^{I(0)}={\cal H}_{00}^I -E_I {\cal N}_{00}^I=0,
\label{eq:HWeq3}
\end{equation}
which gives the trivial solution $E_I={\cal H}_{00}^I/{\cal N}_{00}^I$.
In this way, the structure of the eigenvalue problem is completely
different for $\omega_{\rm rot}\ne 0$, and this is the source of
the discontinuity of the rotational spectra seen in Fig.~\ref{fig:Errotsp}
in the $\omega_{\rm rot}\rightarrow 0$ limit.

\begin{figure}[!ht]
\begin{center}
\includegraphics[width=80mm]{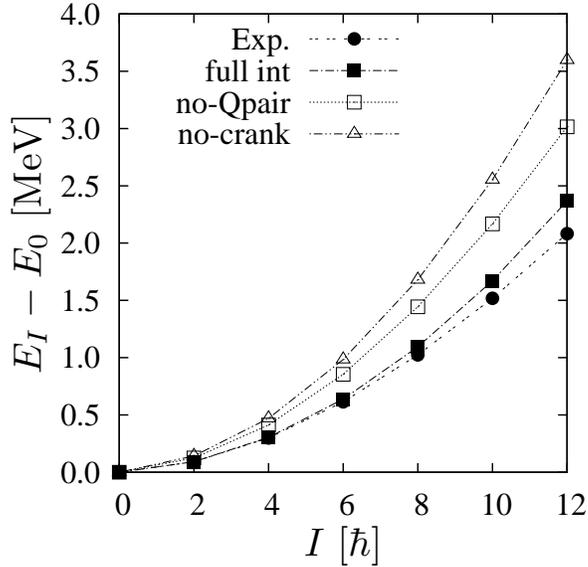}
\vspace*{-5mm}
\caption{
Comparison of the experimental rotational spectra with
the calculated results with various approximations
in $^{164}$Er.
}
\label{fig:ErEI}
\end{center}
\end{figure}

Finally in Fig.~\ref{fig:ErEI},
we compare the experimental rotational spectra
with the results of the quantum number projection calculation.
Here we also included the result of an approximation without cranking,
and that with neglecting the quadrupole pairing ($\lambda=2$) component
in the $H_G$ interaction.  As it is already pointed out
the cranking procedure is important to obtain the correct moment of inertia.
The effect of the quadrupole pairing interaction is also considerable,
and the moment of inertia can be reproduced only within $20\%$ without it.
It should be mentioned that the calculated moment of inertia
for high spin members are underestimated.
In experiment it is known that the moment of inertia
increases as a function of spin;
the amount of increase is about $20\%$ at $I=10$ in $^{164}$Er.
However, the calculated moment of inertia is fairly constant
for high spin members.
The effect of rotation on the mean-field should be included
to obtain the proper amount of increase of the moment of inertia,
which is a future problem.

\subsection{Parity doublet bands in $^{\bf 226}$Th}
\label{sec:Th}

The next example is also a typical rotational spectrum but
with parity violation.
There are several places in nuclear chart, where the static octupole
deformation ($\alpha_{30}$) is expected~\cite{BN96}.
We take a nucleus $^{226}$Th from the actinide region,
which exhibits a nice rotational spectra with alternating parity
based on the parity doublet bands.
The parameters are chosen in the same ways as in the previous
example in $^{164}$Er; especially we took the same value
for the ratio of quadrupole and mono pole pairing force strengths.
The additional deformation parameter is $\alpha_{30}$,
which is chosen to reproduce the splitting of the parity doublet
bands near the band heads.
The resultant parameters are summarized in Table~\ref{tab:Th}.
The value of $\alpha_{30}$ is found to be consistent with
the calculation in Ref.~\cite{BN96}.
The numbers of points for the Gaussian quadrature with respect to
the Euler angles are $N_\alpha=N_\gamma=12$, $N_\beta=60$,
and to the gauge angle $N_\varphi=21$.
As is discussed in the previous section
the Coriolis coupling is necessary to reproduce the moment of inertia
even for the low lying spectra,
so that we use the small cranking frequency $\omega_{\rm rot}=0.01$ MeV
in all the calculations.

\begin{table}[ht]
\begin{center}
\begin{tabular}{ccccccccc}
\hline
$\alpha_{20}$ & $\alpha_{30}$ & $\alpha_{40}$ & $\chi$ [MeV$^{-1}$] &
${\mit\Delta}_{\rm n}$ [MeV] & ${\mit\Delta}_{\rm p}$ [MeV] & 
$g_0^{\rm n}$ [MeV] & $g_0^{\rm p}$ [MeV] & $g_2^\tau/g_0^\tau$ \cr
\hline
0.164\  & 0.075\  & 0.092\  & $1.732\times10^{-4}$ & 0.814 & 0.830 &
0.1140 & 0.1583 & 13.60 \cr
\hline
\end{tabular}
\end{center}
\caption{
The parameters used in the calculation for $^{226}_{\ 90}$Th$_{136}$.
The values of $\chi$ and $g_0^\tau$ are those for the size of
basis $\noscmax=18$.
}
\label{tab:Th}
\end{table}

\begin{figure}[!ht]
\begin{center}
\includegraphics[width=150mm]{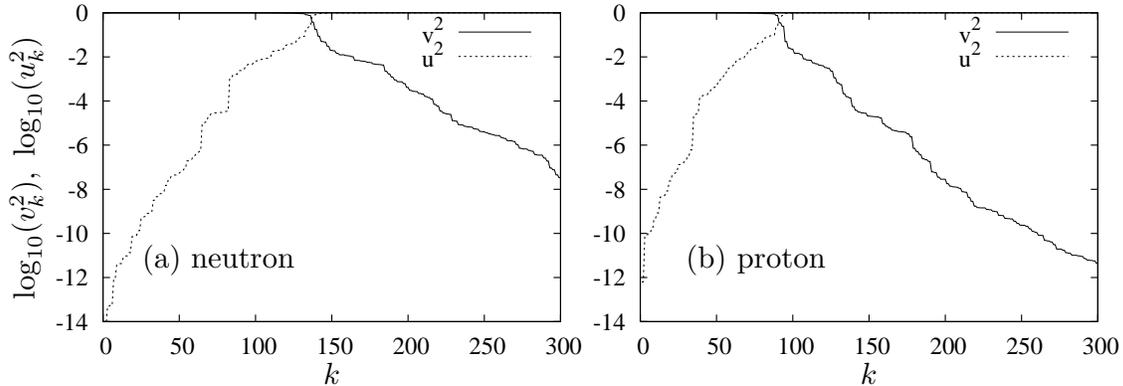}
\vspace*{-5mm}
\caption{
Occupation and empty probabilities $v_k^2$ and $u_k^2=1-v_k^2$
as functions of the number $k$ of the canonical basis for $^{226}$Th;
the log scale is used for the abscissa.
The panel (a) is for neutron and (b) for proton.
}
\label{fig:Thvvuu}
\end{center}
\end{figure}

\begin{figure}[!ht]
\begin{center}
\includegraphics[width=130mm]{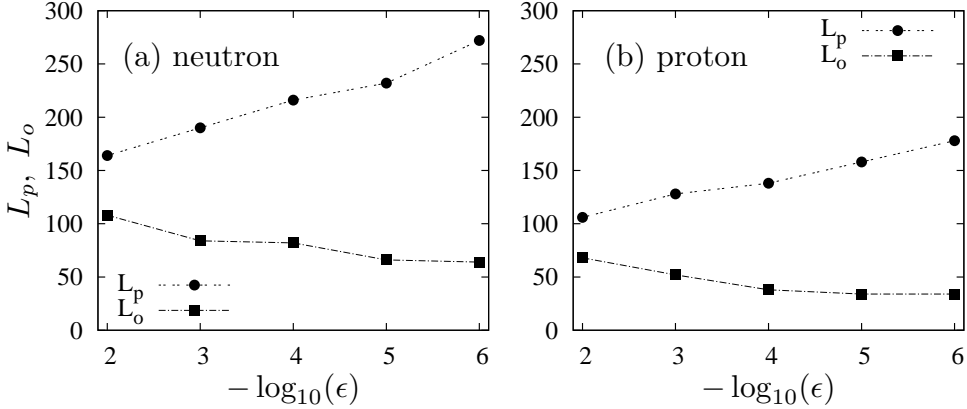}
\vspace*{-5mm}
\caption{
The number of levels in the model space ($P$-space) $L_p$
and the the number of core levels $L_o$ as functions of
the small number $\epsilon$ for $^{226}$Th;
the log scale is used for the ordinate.
The panel (a) is for neutron and (b) for proton.
}
\label{fig:ThepsLp}
\end{center}
\end{figure}

In Figures~\ref{fig:Thvvuu} and~\ref{fig:ThepsLp},
we show the occupation probabilities of the canonical basis and
the dimensions of the truncated model (core) space $L_p$ ($L_o$) for $^{226}$Th,
respectively, as in the previous example.
It is apparent that the pairing correlations are not so strong
that the number of canonical orbits strongly contributing is
a few hundreds and rather small.
The neutron number $N=136$ is relatively large,
and so is the size of the core space,
$L_o \approx 80$ for $\epsilon \approx 10^{-4}$.
Although the quantity $L_p$ is also rather large compared to
the previous case of $^{164}$Er, the difference $L_p-L_o$ is not very
different from that of $^{164}$Er.
Therefore, the method to separate the core contribution in \S\ref{sec:truncphv}
helps to reduce the numerical tasks considerably.
In this way it has been shown that the method of general quantum number
projections developed in \S\ref{sec:formulation} is very efficient
especially when applied to heavier nuclei.

\begin{figure}[!ht]
\begin{center}
\includegraphics[width=130mm]{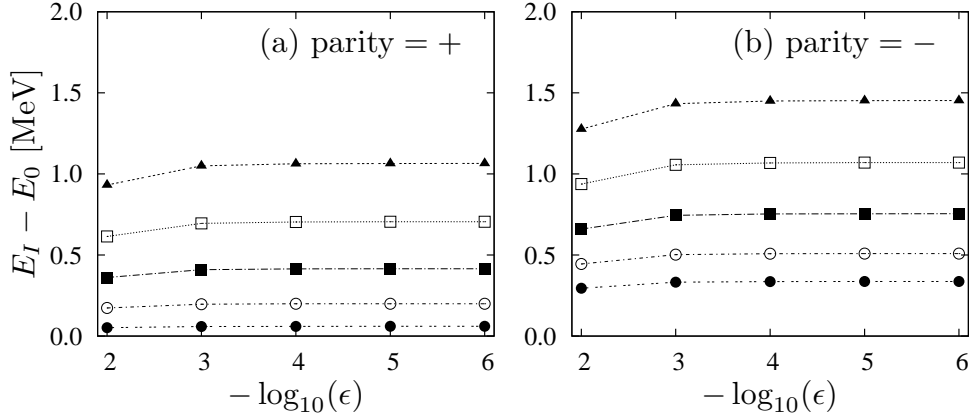}
\vspace*{-5mm}
\caption{
The rotational excitation spectra
calculated by the angular momentum projection for $^{226}$Th
as functions of the cut-off parameter $\epsilon$.
The states from $I=2$ to~8 for $\pi=+$
and those from $I=1$ to~7 for $\pi=-$ are included.
The mean-field state with $\noscmax=18$ and $\hbar\omega_{\rm rot}=0.01$ MeV
is used.
}
\label{fig:Thconvvvl}
\end{center}
\end{figure}

\begin{figure}[!ht]
\begin{center}
\includegraphics[width=150mm]{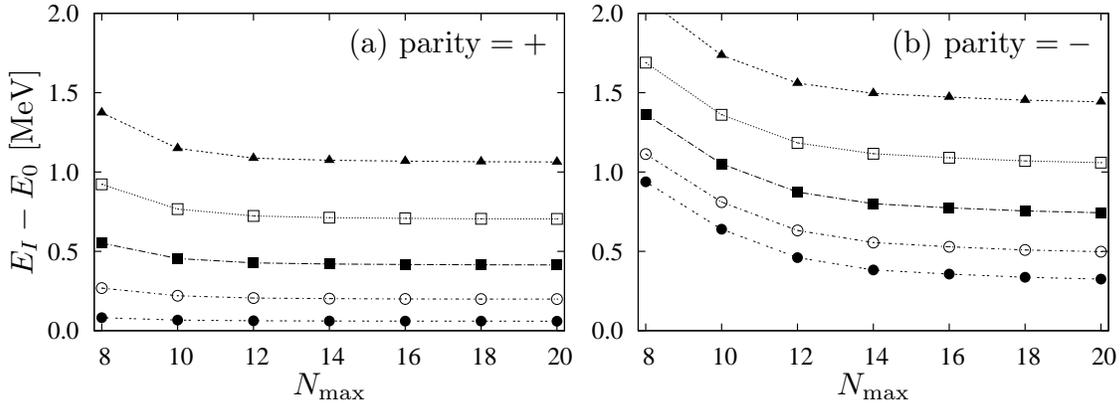}
\vspace*{-5mm}
\caption{
The rotational excitation spectra
calculated by the angular momentum projection for $^{226}$Th
as functions of the size of the spherical harmonic oscillator basis $\noscmax$.
The states from $I=2$ to~8 for $\pi=+$
and those from $I=1$ to~7 for $\pi=-$ are included.
The cranking frequency $\hbar\omega_{\rm rot}=0.01$ MeV is used.
}
\label{fig:ThconvNmax}
\end{center}
\end{figure}

In the case of the static octupole deformation the mean-field states
mix the parity, and the parity projection is necessary.
Because of the signature and time reversal symmetry
in the axially symmetric ground state the even (odd) spin
is only allowed for positive (negative) parity states.
The convergence of the final rotational spectra for each parity
against the cut-off parameter $\epsilon$ is shown in Fig.~\ref{fig:Thconvvvl}.
As in the case of $^{164}$Er the value of $\epsilon \approx 10^{-4}$
is almost enough to attain the stable results for both $\pi=\pm$.
In Fig.~\ref{fig:ThconvNmax} the projected spectra are shown as functions
of the size of the spherical oscillator basis $\noscmax$.
It can be seen that the convergence with respect to $\noscmax$ is
slower for the negative parity states, which have generally
higher excitation energies.
$\noscmax=18$ is almost enough for the positive parity states,
while it may not for the negative parity high spin states.
The small oscillator space like $\noscmax=10$ is dangerous
because the energy of $1^-$ state is overestimated
by more than 200 keV, although the first $2^+$ is almost correct;
the confirmation of the convergence
with respect to the basis size is important.

\begin{figure}[!ht]
\begin{center}
\includegraphics[width=150mm]{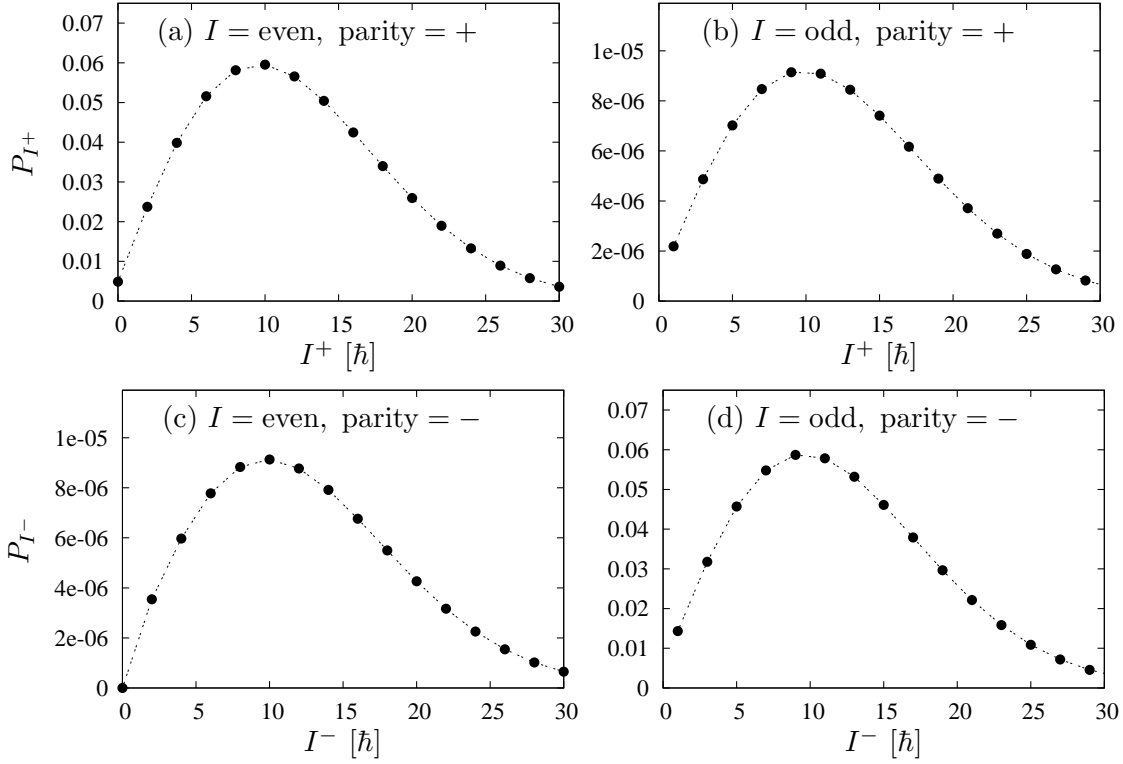}
\vspace*{-5mm}
\caption{
The $I$ distribution of the mean-field state in $^{226}$Th.
The cranking frequency is $\hbar\omega_{\rm rot}=0.01$ MeV.
Even and odd $I$ distributions for $\pi=\pm$ are plotted separately,
because the absolute values are very different.
}
\label{fig:ThIdst}
\end{center}
\end{figure}

In Fig.~\ref{fig:ThIdst} is shown the $I$ distributions of the parity broken
mean-field state in $^{226}$Th.
Only the converged results are included for both parities $\pi=\pm$:
\begin{equation}
 P_{I^\pm}\equiv \sum_{K} \langle\Phi|\hat P^{I}_{KK} \hat P_\pm|\Phi\rangle
 /\langle\Phi|\Phi\rangle.
\end{equation}
The pattern of the distribution is similar to the case of $^{164}$Er,
however, the values of the normal parity components,
$I$=even for $\pi=+$ or $I$=odd for $\pi=-$,
in $^{226}$Th are about half of those in $^{164}$Er.
This is because that the mean-field $|\Phi\rangle$ contains both
$\pi=\pm$ states with almost equal probabilities.
Again the non-normal parity components are much smaller
than the normal parity components because of the cranking procedure
with small frequency $\omega_{\rm rot}=0.01$ MeV.

\begin{figure}[!ht]
\begin{center}
\includegraphics[width=150mm]{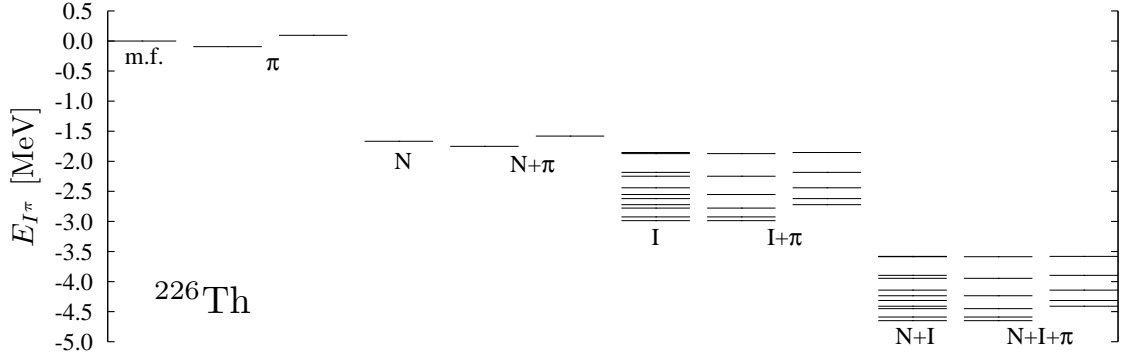}
\vspace*{-5mm}
\caption{
The energy gain by each projection procedure in $^{226}$Th;
the parity ($\pi$), the number ($N$), and the angular momentum ($I$)
projections are done separately as is shown in the figure.
For the results of parity projection, the left side spectra are
of $\pi=+$ and the right side $\pi=-$.
The origin of the energy is that of the mean-field state indicated
by ``m.f.''.
}
\label{fig:ThPNJ}
\end{center}
\end{figure}

The quantum number projection from the symmetry broken mean-field states
are known to be an efficient method to include the correlations
with respect to the collective motions related to the symmetry, e.g.,
the rotational correlations in the case of the angular momentum projection.
In order to show how much correlation energies can be gained in this
particular case, we depict the correlation energies in Fig.~\ref{fig:ThPNJ}.
The energy gain by the parity projection is not large and
less than a few hundred keV,
while those by the number projection
(adding the contributions from both neutron and proton) and
the angular momentum projection are about 1.5 MeV and 3.0 MeV, respectively,
and the total amount is $-4.35$ MeV in this calculation.

\begin{figure}[!ht]
\begin{center}
\includegraphics[width=80mm]{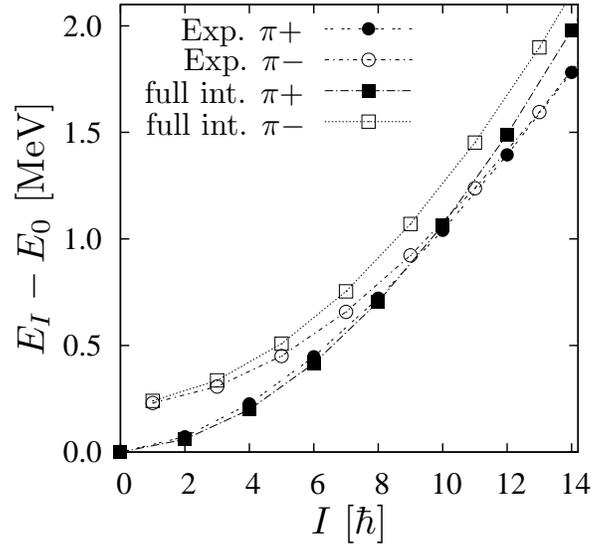}
\vspace*{-5mm}
\caption{
Comparison of the experimental rotational spectra with
the calculated results in $^{226}$Th.
}
\label{fig:ThEI}
\end{center}
\end{figure}

Finally we compare the $\pi=\pm$ rotational spectra with
experimental data in Fig.~\ref{fig:ThEI}.
In this calculation one cranked mean-field with $\omega_{\rm rot}=0.01$ MeV
is used to generate all the states shown in the figure.
As is similar to the case of $^{164}$Er, the experimental moment of
inertia increases as a function of spin.  The calculated inertia also
increases slightly but far not enough to account for the experimentally
observed trend.
Especially, the degeneracy of the negative and positive parity band
becomes better and better for $I \ge 10$, which is not well reproduced
in the calculation, where the moment of inertia of the high spin part
of the negative parity band is too small.
The present investigation is the simplest in the sense that only
the one intrinsic state is used for all the spins.
We need to improve the description for the high spin states.

\section{Summary}
\label{sec:summary}

In this paper, we have developed an efficient technique to perform
calculations of the quantum number projections from
the most general HFB type state; the the configuration mixing can be
done additionally with the same technique if necessary.
The use of the HFB type mean-field, i.e., including the effect
of the pairing correlation, generally requires
a large model space in realistic situations.
Our basic strategy is to transform the original basis into the canonical basis,
and to discard the orbits with small occupation probabilities.
We have shown that the truncation scheme works very well,
i.e., the convergence is very rapid and the number orbits which should
be included in the calculation is reduced more than an order of magnitude.
With this truncation scheme, it has been demonstrated that
the angular momentum projection calculation
with the spherical oscillator shell more than $\noscmax=20$ is possible.

Another characteristic feature of our approach is that the Thouless amplitude
with respect to a Slater determinant is utilized for the projection and
configuration mixing calculations.  In this way the calculations are
divided into the part related to the Slater determinant and the part
taking into account the pairing correlation.
With this technique, the Thouless
amplitude never diverges for the case of small pairing correlations,
or for the case where the blocked levels exist, e.g., for the odd nuclei,
and calculation can be performed reliably.
Moreover, this makes it possible to exclude
the core contributions for the pairing correlation and to further reduce
the dimension of the various matrix operations.
It has been demonstrated that this elimination of the core contribution
is especially effective for heavy nuclei like in the actinide region.

As for the test calculations, we have set up
the schematic separable interactions suitable for the Woods-Saxon potential,
and performed realistic calculations for a rare earth and an actinide nucleus.
The quadrupole pairing interaction is included for the pairing channel.
The number as well as the angular momentum projections
have been done at the same time.
In this paper we have used only one mean-field state (no configuration mixing),
and tried to describe the typical rotational spectra
with good agreements for the low spin states.
It has been shown that the cranking with small frequency
in the mean-field is very important
to reproduce the experimental value of moment of inertia.
In the case of the actinide nucleus $^{226}$Th,
which is believed to have a pear shape,
the parity projection has been also done simultaneously.
With a reasonable octupole deformation parameter, the excitation energy
of the negative parity band head can be reproduced.
However, the calculated moment of inertia with one intrinsic state
is almost constants within the rotational band and does not well describe
its gradual increase at higher spin states observed in experiments.
Therefore, the effect of change of the mean-field or of the configuration
mixing is necessary to obtain a better description of higher spin states,
which is an important future problem.

\section{Acknowledgements}
This work is supported
by Grant-in-Aid for Scientific Research~(C)
No.~22540285 from Japan Society for the Promotion of Science.

\section{Appendix}
\label{sec:append}

In this Appendix, we show that the model space truncation scheme,
which was first introduced in the Appendix of Ref.~\cite{BFH90}
in terms of the $(U,V)$ amplitudes, can be naturally derived
from our formulation in \S\ref{sec:formulation}.
Note that this can be done without any problem
when the truncated dimensions of the left and right states
are the same, i.e., $L_p=L_{p'}$
(see also \cite{VHB00} and the Appendix of Ref.~\cite{YM09}).
We assume it throughout in this Appendix.

The basic quantities that should be evaluated in the projection
are the overlap of the transformation operator
$\langle \Phi|\hat D|\Phi'\rangle$, and the associated contractions
of the creation and annihilation operators in Eq.~(\ref{eq:Dcontc}),
which are calculated through those with respect to the canonical basis
in Eq.~(\ref{eq:bDcont}).
Here we assume that the HFB type states $|\Phi\rangle$ and  $|\Phi'\rangle$
are normalized.  Then by using the normalization constants
in Eq.~(\ref{eq:normphase}) and the $(\bar U,\bar V)$ amplitudes
in the canonical basis in Eq.~(\ref{eq:canWU}),
the overlap in Eq.~(\ref{eq:ovlZDp}) can be calculated as
\begin{eqnarray}
 \langle \Phi |\hat D |\Phi' \rangle
 &=& e^{i(\theta'_1-\theta^{}_1)}\langle |\hat D|\rangle
  \left( \det \left(\bar U_{pp}^\dagger \bar U'_{p'p'}\right)^*
  \det \left[1+Z_{pp}^\dagger Z'_{D{pp}}\right]\right)^{1/2} \cr
 &=& e^{i(\theta'_1-\theta^{}_1)}\langle |\hat D|\rangle
  \left( \det \left(\bar U_{pp}^\dagger \bar U'_{p'p'}\right)^*
  \det \left[
    \tilde D_{pp^\prime}^{-T}+Z_{pp}^\dagger \tilde D_{pp'} Z'_{p'p'} \right]
     \det \bigl(\tilde D_{pp'}^T\bigr)\right)^{1/2} \cr
 &=& e^{i(\theta'_1-\theta^{}_1)}\langle |\hat D|\rangle
  \left( \det \tilde D_{pp'} \det A_{pp'} \right)^{1/2},
\label{eq:AovlUV}
\end{eqnarray} 
where $\theta'_1$ and $\theta_1$ are the arbitrarily chosen phases
for the states $|\Phi\rangle$ and $|\Phi' \rangle$, respectively.
In Eq.~(\ref{eq:AovlUV}) are used the definitions of the $Z$ amplitude
in Eq.~(\ref{eq:ZUVp}) and of the $Z'_D$ in Eq.~(\ref{eq:trZDpq}),
and the new matrix $A_{pp'}$ is introduced by
\begin{equation}
 A_{pp'}\equiv
 \bar U_{pp}^T \tilde D_{pp^\prime}^{-T} \bar U'^*_{p'p'}
   +\bar V_{pp}^T \tilde D_{pp^\prime } \bar V'^*_{p'p'}
\label{eq:defA}.
\end{equation} 
Note that because of $L_p=L_{p'}$ the inverse of the matrix
$\tilde D_{pp'}$ is well defined.
In this way, the overlap can be calculated within the $P$ space.
However, the sign problem of the square root remains
in this form~(\ref{eq:AovlUV}).

The contractions in Eq.~(\ref{eq:bDcont}) can be calculated
in terms of the $(\bar U,\bar V)$ amplitudes in the same way:
\begin{eqnarray}
 \rho^{(b)}_{Dpp} &=&
 Z'_{Dpp} \left[1+Z_{pp}^\dagger Z'_{Dpp}\right]^{-1} Z_{pp}^\dagger
 = \tilde D_{pp'} Z'_{p'p'}
 \left[\tilde D_{pp'}^{-T}+Z_{pp}^\dagger \tilde D_{pp'} Z'_{p'p'} \right]^{-1}
 Z_{pp}^\dagger
\nonumber\\
 &=& \tilde D_{pp'} \bar V'^*_{p'p'} A_{pp'}^{-1} \bar V_{pp}^T,
\\
 \kappa^{(b)}_{Dpp} &=&
 Z'_{Dpp} \left[1+Z_{pp}^\dagger Z'_{Dpp}\right]^{-1}
 = \tilde D_{pp'} Z'_{p'p'}
 \left[\tilde D_{pp'}^{-T}+Z_{pp}^\dagger \tilde D_{pp'} Z'_{p'p'} \right]^{-1}
\nonumber\\
 &=& \tilde D_{pp'} \bar V'^*_{p'p'} A_{pp'}^{-1} \bar U_{pp}^T,
\\
 \bar \kappa^{(b)}_{Dpp} &=&
 \left[1+Z_{pp}^\dagger Z'_{Dpp} \right]^{-1} Z_{pp}^\dagger
 =\tilde D_{pp'}^{-T} 
 \left[\tilde D_{pp'}^{-T}+Z_{pp}^\dagger \tilde D_{pp'} Z'_{p'p'} \right]^{-1}
 Z_{pp}^\dagger
\nonumber\\
 &=& \tilde D_{pp'}^{-T}  \bar U'^*_{p'p'} A_{pp'}^{-1} \bar V_{pp}^T,
\end{eqnarray}
which are calculated within the $P$ space.
By using Eq.~(\ref{eq:cbDcontrn}) we finally obtain
\begin{eqnarray}
 \rho^{(c)}_D &=&
 D W'_{p'}(\bar V'^*_{p'p'} A_{pp'}^{-1} \bar V_{pp}^T)W_p^\dagger,
\\
 \kappa^{(c)}_D &=&
 D W'_{p'}(\bar V'^*_{p'p'} A_{pp'}^{-1} \bar U_{pp}^T)
 \tilde D_{pp'}^{-T} W'^T_{p'} D^T,
\\
 \bar \kappa^{(c)}_D &=&
 W^*_p \tilde D_{pp'}^{-T}(\bar U'^*_{p'p'} A_{pp'}^{-1} \bar V_{pp}^T)
 W_p^\dagger.
\end{eqnarray}
Thus all the corresponding quantities can be calculated within
the truncated canonical basis in terms of
the $(\bar U,\bar V)$ amplitudes in place of the Thouless amplitudes.

In the general case $L_p \ne L_{p'}$, the definitions of the inverse
matrices $\tilde D_{pp'}^{-1}$ and $A_{pp'}^{-1}$ are ambiguous,
and more careful analysis is necessary.
There is no such difficulty in our formulation in terms of
the Thouless amplitudes.


\end{document}